\documentclass[12pt,a4paper]{article}
\usepackage{fullpage}

\usepackage[utf8]{inputenc}
\usepackage{array}
\usepackage{graphicx}
\usepackage{verbatim}
\usepackage{amsmath}
\usepackage{amsfonts}
\usepackage{amssymb}
\usepackage{caption}
\usepackage{cite}
\usepackage[breaklinks]{hyperref}
\usepackage{subcaption}
\usepackage{float}
\usepackage{authblk}
\usepackage{multirow}
\usepackage{rotating}
\usepackage{section}
\usepackage{cancel}
\usepackage{tikz}
\usepackage[compat=1.1.1]{tikz-feynman}
\usepackage{feynmp-auto}

\makeatletter
\newcommand\extralabel[2]{{\edef\@currentlabel{\@currentlabel#2}\label{#1}}}
\makeatother 

\begin{document}
	\title{Neutrino Model in Left-Right Symmetric Linear Seesaw Augmented with $A_4$ Modular Group}
	
	\author{Raktima Kalita\textsuperscript{1,}\thanks{$^*$email: phy2091007\textunderscore raktima@cottonuniversity.ac.in}  and Mahadev Patgiri\textsuperscript{1,}\thanks{$^\dagger$email: mahadevpatgiri@cottonuniversity.ac.in}}
	\affil{\textsuperscript{1}Department of Physics, Cotton University, Guwahati, India}		
	\maketitle	
	
\begin{abstract}
 \par In this work, we have implemented $A_4$ modular symmetry in the left-right symmetric linear seesaw model. Interestingly, such modular symmetry restricts the proliferation of flavon fields, and as a result, the predictibility of the model is enhanced. The fermion sector of the model comprises of quarks, leptons and a sterile fermion in each generation, while the scalar sector consists of Higgs doublets and bidoublets. We investigate numerically various Yukawa coupling co-efficients, the neutrino masses and mixing parameters in our intended model and predictions become consistent with $3\sigma$ range of current neutrino oscillation data. We also studied the non-unitarity, effects on lepton flavor violation in our model and evolution of lepton asymmetry to explain the current baryon asymmetry of the universe.
\end{abstract}  

\section{Introduction}
\par The neutrino oscillations confirmed in 1998 by the Super-Kamiokande experiment was a watershed in the history of neutrino physics. This indicated that neutrinos are massive and mixing. Interestingly, neutrinos have tiny masses in sub-eV scale, while two mixing angles are large. In Standard Model (SM) of particle physics, neutrinos are massless, so massive neutrinos are the physics beyond SM (BSM). In the standard parametrization of neutrino mass matrix, the neutrino oscillation parameters, viz., three neutrino mixing angles $(\theta_{ij})$ and two mass squared differences $(\Delta m_{ij}^2)$ have been determined in experiments with a good precision and accuracy. However, some important parameters like the Dirac CP violating phase $(\delta_{CP})${\cite{acciarri2016long,t2k2020constraint}}, octant degeneracy of atmospheric mixing angle $(\theta_{23})$, neutrino mass hierarchy and absolute neutrino mass scale are yet to determine. As theoretical attempts to understand the experimental results about neutrinos, a number of neutrino mass models, namely, seesaw mechanism\cite{minkowski1977mu,mohapatra1980neutrino,schechter1980neutrino} with its varieties, radiative mass generation models\cite{zee1980theory,babu1988model} etc. have been proposed in literature. The existence of sterile neutrinos is one of the key features of BSM scenarios. SM gauge singlets, which are right-handed neutrinos that interact with standard active neutrinos through Yukawa interactions, are generally thought of as sterile neutrinos. The sub-eV scale light neutrinos in the canonical seesaw framework are explained by assuming that the mass of right-handed neutrinos is of the order of $10^{15}$ GeV, which is beyond the scope of present and forthcoming collider experiments. However, some low-scale mechanisms, such as the extended seesaw\cite{mohapatra2007theory}, linear seesaw\cite{malinsky2005supersymmetric,dib2014neutrinos}, inverse seesaw\cite{gonzalez1989fast,hernandez2019viable,brdar2019low} etc. require neutrino mass to be in the TeV scale to make them experimentally testable.

\par It is noted that group symmetries play a crucial role in particle physics. In the literature, wide varieties of neutrino mass models using non-abelian discrete flavor symmetries like $S_4$, $A_4$, $A_5$ etc. are available \cite{brahmachari20084,mukherjee2016neutrino,di2019neutrino,ma2006neutrino,chakraborty2020predictive} . The detailed analysis of these mass models and also their phenomenological implications have been attempted. In this platform, the models require extra flavon fields which are generally SM singlets in the process of their symmetry realization. These flavons play a vital role in understanding of the observed pattern of neutrino masses and mixings by virtue of their particular vacuum alignment\cite{pakvasa1978discrete}. However, the proliferation of flavon fields make the model less predictive. Though flavor symmetry is used to constrain mixing angles but except for few scenarios, the neutrino masses remain undetermined. To get rid of flavon fields, an approach of modular invariance \cite{feruglio2019neutrino} has slowly been attracting attentions of a few authors working in neutrino mass models. One of the striking features of this approach is that it may not require flavon fields other than the modulus, which enhances the power of predictability of the model\cite{xing2020flavor}.
\par In the present work, we deal a left-right symmetric linear seesaw model \cite{sahu2020$a_4$} with modular $A_4$ symmetry instead of a linear seesaw model with modular $A_4$ symmetry and global symmetry \cite{behera2022implications,nomura2022linear}. We will construct the desired neutrino mass matrix in the proposed framework and perform a systematic numerical analysis to predict various neutrino parameters which are compatible with the current neutrino data. 

\par The organization of this paper is as follows. In Sec. \ref{sec_2}, the framework of the left-right symmetric linear seesaw mechanism has been presented. In Sec. \ref{sec_3}, there is a brief overview of the modular symmetry and its implications in particle physics. In Sec. \ref{sec_4}, we have constructed our model of the neutrino mass matrix using a linear seesaw mechanism with modular $A_4$ symmetry. In Sec. \ref{sec_5}, a detailed numerical analysis has been performed and examined the compatibility of our model predictions of neutrino observables with the current neutrino data.  In Sec. \ref{sec_6} and Sec. \ref{sec_7}, we have discussed lepton flavor violating processes and leptogenesis in context of our model. In Sec. \ref{sec_8}, we briefly discuss on the collider implications for the present model. Finally, in Sec. \ref{sec_9} we have presented the discussion and conclusion of the results of our proposed work.
\section{Left-right symmetric linear seesaw model} \label{sec_2}
\par The left-right symmetric theory has advanced significantly, providing answers to a number of additional questions including the small mass of neutrino and dark matter sector left unaccounted for by the Standard Model. It was first put forth to explain the origin of parity violation in low-energy weak interactions. The left-right symmetric model (LRSM), introduced by Pati and Salam, is an extension of the Standard Model(SM) of particle physics with the following gauge group\cite{grimus1993introduction}
\begin{equation}
	SU(3)_C\times SU(2)_L\times SU(2)_R\times U(1)_{B-L} \label{eqn_1}
\end{equation}
where $B-L$ is the difference between baryon number and lepton number.
\par The quark and lepton contents of the model are given by
\begin{equation}
	\begin{aligned}
		q_L=\begin{pmatrix} u_L\\ d_L \end{pmatrix}\sim(3,2,1,\frac{1}{3}), \;  q_R=\begin{pmatrix} u_R\\ d_R \end{pmatrix}\sim(3,1,2,\frac{1}{3})\\ \label{eqn_2}
	\end{aligned}
\end{equation}
\begin{equation}
	\begin{aligned}
		l_L=\begin{pmatrix} \nu_L\\ e_L \end{pmatrix}\sim(1,2,1,-1), \;  l_R=\begin{pmatrix} \nu_R\\ e_R \end{pmatrix}\sim(1,1,2,-1)\\ \label{eqn_3}
	\end{aligned}
\end{equation}
\par There are the following two Higgs doublets and a bidoublet in the scalar sector of LRSM.
\begin{equation}
	\begin{aligned}
		H_L=\begin{pmatrix} h_L^{-}\\ h_L^{0} \end{pmatrix}\sim(1,2,1,-1), \;  H_R=\begin{pmatrix} h_R^{+}\\ h_R^{0} \end{pmatrix}\sim(1,1,2,-1)\\ \label{eqn_4}
	\end{aligned}
\end{equation}
\begin{equation}
	\Phi=\begin{pmatrix} \phi_1^{0} & \phi_2^{+}\\ \phi_1^{-} & \phi_2^{0}\end{pmatrix}\sim (1,2,2,0) \label{eqn_5}
\end{equation}
\par  Again the left-right symmetric linear seesaw requires a neutral gauge singlet fermion $S$ in order to produce neutrino masses. Thus, we obtain the Lagrangian for the leptonic sector as follows:
\begin{equation}
	-\mathcal{L}_{lepton}=\bar{l}_LY\Phi l_R+\bar{l}_LY\tilde{\Phi} l_R+Y\bar{l}_LH_LS+Y\bar{l}_RH_RS+h.c. \label{eqn_6}
\end{equation}
where $\tilde{\Phi}=\sigma_2\Phi^\ast\sigma_2$ and $\sigma_2$ is the second Pauli matrix.
\par After acquiring vevs, the scalars $H_L$, $H_R$ and $\Phi$ can be written as
\begin{equation}
	\begin{aligned}
		\langle H_L\rangle=\begin{pmatrix} v_L\\ 0 \end{pmatrix}, \;  
		\langle H_R\rangle=\begin{pmatrix} v_R\\ 0 \end{pmatrix}, \; 
		\langle\Phi\rangle=\begin{pmatrix} v_1 & 0\\ 0 & v_2\end{pmatrix}\\ \label{eqn_7}
	\end{aligned}
\end{equation}
\par Thus, the effective $9\times9$ neutrino mass matrix in the basis $(\nu_L,\nu_{R}^c,S)$ becomes
\begin{equation}
	M_\nu=\begin{pmatrix}
		0 & m_{D} & m_{LS}\\ m_{D}^T & 0 & m_{RS} \\ m_{LS}^T & m_{RS}^T & 0	\end{pmatrix} \label{eqn_8}
\end{equation}
\par In the above matrix, the $\mu$ term is absent because of the introduction of $A_4$ symmetry. With $m_{LS}\ll m_{D}< m_{RS}$, the light neutrino mass formula is given by
\begin{equation} 
	m_\nu=m_{D}m_{RS}^{-1}m_{LS}^T+ transpose \label{eqn_9}
\end{equation}
\par In addition, the other important neutrino parameters, viz.,the effective neutrino mass, $m_{ee}$ and the Jarlskog invariant, $J_{CP}$ which are the measure of the neutrinoless double beta decay rate and strength of CP violation respectively, are given by
\begin{equation}
    |m_{ee}|=|m_1\cos^2{\theta_{12}}\cos^2{\theta_{13}}+m_2\sin^2{\theta_{12}}\cos^2{\theta_{13}}e^{i\alpha_{21}}+m_3\sin^2{\theta_{13}}e^{i(\alpha_{31}-2\delta_{CP})}| \label{eqn_10}
\end{equation}
\begin{equation}
	\begin{split}
	J_{CP}&=\text{Im}[U_{e1}U_{\mu2}U_{e2}^*U_{\mu1}^*] \\ &=\sin{\theta_{23}}\cos{\theta_{23}}\sin{\theta_{12}}\cos{\theta_{12}}\sin{\theta_{13}}\cos^2{\theta_{13}}\sin{\delta_{CP}} \label{eqn_11}
	\end{split}
\end{equation}
\par The detection of the rare events like neutrinoless double beta decay, a lepton number violation process\cite{schechter1982neutrino} is very challenging and will confirm the Majorana nature of neutrinos. In this context, the ongoing experiments, viz., GERDA and CUORE are having the sensitivity limits for $|m_{ee}|$ to $(102-213)$ meV\cite{agostini2019probing} and $(110-520)$ meV\cite{alduino2018first} respectively. The future generation experiments, LEGEND-200 and KamLAND-Zen have their capabilities to provide the sensitivities for $|m_{ee}|$ up to $(35-73)$ meV\cite{giuliani2019double} and $(61-165)$ meV\cite{gando2016search} respectively.

\section{Formalism of Modular Symmetry} \label{sec_3}
\par The concept of modular transformation and its subgroup symmetry was first introduced in a superstring theory to explain the process of compactification of exra dimensions of space. The two dimensional torus can be constructed as $R^2$ divided by a two dimensional lattice $\Lambda$, which is spanned by the vectors $\left(\alpha_{1}, \alpha_{2}\right)= \left(2\pi R, 2\pi R\tau\right)$ with R being real and $\tau$ a complex {\it modulus} parameter\cite{feruglio2019neutrino,de2012finite,kobayashi2018neutrino,kobayashi2018modular}. Now the same lattice can also be described in another basis,\\
\begin{equation}
	\begin{pmatrix}
		\alpha^{\prime}_{2}\\
		\alpha^{\prime}_{1}
	\end{pmatrix}= \begin{pmatrix}
		a & b\\
		c & d
	\end{pmatrix} \begin{pmatrix}
		\alpha_{2}\\
		\alpha_{1}
	\end{pmatrix} \label{eqn_12}
\end{equation}
where $a$, $b$, $c$, and $d$ are integers and satisfy $ad-bc=1$, the modulus parameter $\tau$ transforms as\cite{xing2020flavor}
\begin{equation}
\tau=\frac{\alpha_{2}}{\alpha_{1}}\rightarrow \tau^{\prime}=\frac{\alpha^{\prime}_{2}}{\alpha^{\prime}_{1}}=\frac{a\tau+b}{c\tau+d} \label{eqn_13}
\end{equation}
This modular transformation is generated by $S$:$\tau\rightarrow -1/\tau$ and $T\rightarrow\tau +1$ which satisfy $S^{2}=I$ and $\left(ST\right)^{3}=I$. If $T^{N}=I$ is required, one is left with the finite subgroups $\Gamma_{N}$ which are isomorphic to some even permutation groups, particularly, $\Gamma_{2}\simeq S_{3}$, $\Gamma_{3}\simeq A_{4}$, $\Gamma_{4}\simeq S_{4}$, $\Gamma_{5}\simeq A_{5}$ \cite{de2012finite}. The holomorphic functions transforming as $f(\tau)\rightarrow(c\tau +d)^{k}f(\tau)$ under modular transformation in Eqn. \eqref{eqn_13} are called the modular forms of weight $k$.
It becomes clear that the modular groups $\Gamma_{2,...5}$ can find some interesting applications in understanding flavor structures of quarks and leptons. The advantage of a modular symmetry over a usual discrete flavor symmetry is that the former dictates the Yukawa-like couplings to be the functions of the modulus parameter $\tau$ and transform nontrivially under $\Gamma_{N}$, whereas the latter only acts on the fermion and scalar fields. To achieve the dependence of the Yukawa-like couplings on $\tau$, it has been found that the Dedekind $\eta$-function in the following, is a good example of this kind\cite{feruglio2019neutrino}.
\begin{align}
		\eta(\tau)=q^\frac{1}{24}\prod_{n=1}^\infty(1-q^n) \label{eqn_14}\\  q\equiv e^{2\pi i \tau} \label{eqn_15}
\end{align} 
This satisfies

\begin{align}
		\eta(-1/\tau)=\eta(\tau) \sqrt{-i\tau} \qquad \eta (\tau +1)= \eta(\tau) e^{(i\pi/12)} \label{eqn_16}
\end{align} 

corresponding to aforementioned transformations.

There are some effective neutrino models of modular symmetry investigated in literature which have attracted the attention of many researchers because of its appealing feature of not requiring flavon fields in symmetry realization apart from a complex modulus, $\tau$. The flavor symmetry is broken when $\tau$ acquires vacuum expectation value\cite{king2020fermion}, thus a mechanism is required to fix this modulus $\tau$. Now we outline briefly how modular symmetry is implemented. We consider the modular group, $\Gamma(N)$ which when operates on the complex variable $\tau$ belonging to the upper-half of the complex plane i.e., Im$(\tau)>0$, it transforms as the following \cite{feruglio2019neutrino}:
\begin{equation}
	\gamma(\tau)=\frac{a\tau+b}{c\tau+d} \label{eqn_17}
\end{equation}
where \textit{a,b,c,d} are integers forming a matrix
\begin{equation}
	\begin{aligned}
		\gamma=\begin{pmatrix} a & b\\ c& d \end{pmatrix} \label{eqn_18}
	\end{aligned}
\end{equation} 
with determinant, $ad-bc=1$.
\par The modular group is always isomorphic to the projective special linear group $ PSL(2,Z)=SL(2,Z)/Z_2 $, where $SL(2,Z)$ is a $2\times2$ matrix with determinant unity. The elements $S$ and $T$ with the condition $S^2=ST^3=I$ can generate the modular group  and their matrix representations are in the following:
\begin{equation}
	\begin{aligned}
		S=\begin{pmatrix} 0 & 1 \\ -1 & 0 \end{pmatrix}, \; T=\begin{pmatrix} 1 & 1 \\ 0 & 1 \end{pmatrix} \label{eqn_19}
	\end{aligned}
\end{equation}
Such generators when operate on $\tau$ yield the following transformations:
\begin{equation}
	S:\tau \rightarrow -\frac{1}{\tau}, \; T:\tau \rightarrow \tau + 1 \label{eqn_20}
\end{equation}
\par It is interesting that the Yukawa couplings in the framework of modular symmetry can cast as the functions of the complex modulus $\tau$. With the choice of the modular group, the level and weight of modular symmetry determine the number of modular forms. The number of modular forms related to the non-abelian discrete symmetry group to which it is isomorphic is shown in the table below\cite{feruglio2019neutrino}.
\begin{center}
	\begin{table}[H]
		\caption{No. of modular forms corresponding to weight 2k.}
		\centering
		\begin{tabular}{|c|c|c|}
		\hline
		N & No. of modular forms & $\Gamma(N)$ \\
		\hline
		2 & k+1 & $S_3$ \\
		\hline
		3 & 2k+1 & $A_4$ \\
		\hline
		4 & 4k+1 & $S_4$ \\
		\hline
		5 & 10k+1 & $A_5$ \\
		\hline
		6 & 12k & \\ 
		\hline
		7 & 28k-2 & \\
		\hline
	    \end{tabular}
    \end{table}
\end{center}
\par In our work, we shall use $\Gamma(3)$, which is isomorphic to the discrete symmetry group $A_4$. The three linearly independent modular forms of weight 2 and level-3 are given by
\begin{align}
		Y_1(\tau)=\frac{i}{2\pi}\Biggr[\frac{\eta^\prime(\frac{\tau}{3})}{\eta(\frac{\tau}{3})}+\frac{\eta^\prime(\frac{\tau+1}{3})}{\eta(\frac{\tau+1}{3})}+\frac{\eta^\prime(\frac{\tau+2}{3})}{\eta(\frac{\tau+2}{3})}-\frac{27\eta^\prime(3\tau)}{\eta(3\tau)}\Biggr] \\
		Y_2(\tau)=\frac{-i}{\pi}\Biggr[\frac{\eta^\prime(\frac{\tau}{3})}{\eta(\frac{\tau}{3})}+\omega^2\frac{\eta^\prime(\frac{\tau+1}{3})}{\eta(\frac{\tau+1}{3})}+\omega\frac{\eta^\prime(\frac{\tau+2}{3})}{\eta(\frac{\tau+2}{3})}\Biggr] \\
		Y_3(\tau)=\frac{-i}{\pi}\Biggr[\frac{\eta^\prime(\frac{\tau}{3})}{\eta(\frac{\tau}{3})}+\omega\frac{\eta^\prime(\frac{\tau+1}{3})}{\eta(\frac{\tau+1}{3})}+\omega^2\frac{\eta^\prime(\frac{\tau+2}{3})}{\eta(\frac{\tau+2}{3})}\Biggr] \\ \label{21}
\end{align}
where $\eta(\tau)$ is the Dedekind eta-function, defined in Eqn. \eqref{eqn_14}
\section{Model Framework} \label{sec_4}
\par Here, we briefly discuss about the framework for left-right symmetric model with linear seesaw mechanism. In this framework, we have introduced $A_4$ modular group which minimizes the use of multiple flavon fields. Since modular symmetry has been taken into consideration, we provide the particles with modular weights while keeping in mind that though matter multiplets corresponding to the model may have negative modular weights, but we cannot provide negative weights to the modular forms. These weights are assigned in such a way that the sum of the modular weights in each term of the Lagrangian equals zero. Table \ref{tab_2} provides the particle contents and group charges for the model. The group transformation of the Yukawa couplings is non-trivial in this case. Table \ref{tab_3} lists the group charge and modular weight attributed to the Yukawa couplings.
\begin{table}[H]
	\begin{center}
		\caption{Particle contents and group charges for the model.}
		\begin{tabular}{|c|c|c|c|c|c|c|}
			\hline
			Fields & $SU(3)_c$ & $SU(2)_L$ & $SU(2)_R$ & $B-L$ & $A_4$ & $k_I$ \\
			\hline
			$l_{{L}_{1,2,3}}$ & $1$ & $2$ & $1$ & $-1$ & $1,1'',1'$ & $-1$ \\
			$l_R$ & $1$ & $1$ & $2$ & $-1$ & $3$ & $-1$ \\
			$S$ & $1$ & $1$ & $1$ & $0$ & $3$ & $-1$ \\
			\hline
			$\Phi$ & $1$ & $2$ & $2$ & $0$ & $1$ & $0$ \\
			$H_L$ & $1$ & $2$ & $1$ & $-1$ & $1$ & $0$ \\
			$H_R$ & $1$ & $1$ & $2$ & $-1$ & $1$ & $0$ \\ 
			\hline
		\end{tabular}
  \label{tab_2}
	\end{center}
\end{table}
\begin{table}[H]
	\begin{center}
		\caption{Modular weight attributed to the Yukawa couplings.}
		\begin{tabular}{|c|c|c|}
			\hline
			Yukawa coupling & $A_4$ & $k_I$ \\
			\hline
			$Y$ & $3$ & $2$ \\
			\hline
		\end{tabular}
  \label{tab_3}
	\end{center}
\end{table}
\subsection{Dirac mass term for light neutrinos}
\par In this model, the left-handed lepton doublets $(l_{L_{e}},l_{L_{\mu}},l_{L_{\tau}})$ are transformed as singlet $(1,1^{''},1^{'})$ and the right-handed lepton $l_R$ transformed as triplet under $A_4$ modular symmetry. The modular weight assigned to them is $-1$. The scalar field, i.e., Higgs bidoublet is transformed as singlet under $A_4$ modular group. Similarly. the yukawa couplings $Y=(Y_1,Y_2,Y_3)$ transforms as triplet under this symmetry and the modular weight is $2$.
\par Now, the Lagrangian for the Dirac mass term is given by
\begin{equation}
	\mathcal{L}_{D}=\bar{l}_LY\Phi l_R+\bar{l}_LY\tilde{\Phi}l_R
\end{equation}
\par Therefore, the mass matrix is given by
\begin{equation}
	m_{D}=v_{D}
	\begin{pmatrix}
		Y_1 & Y_3 & Y_2\\
		Y_2 & Y_1 & Y_3\\
		Y_3 & Y_2 & Y_1\\
	\end{pmatrix}
\end{equation}
where $v_{D}$ is the vev for the Higgs bidoublet.
\subsection{Mixing between $\nu_{L}$ and $S$}
\par Here, the additional left-right gauge singlet neutral fermion $S$ transforms as triplet under $A_4$ symmetry. The modular weight assigned to it is $-1$. Also, the scalar field $H_L$ transformed as singlet under the given symmetry.
\par The Lagrangian for this mixing can be written as
\begin{equation}
	\mathcal{L}_{LS}=\alpha_{LS}(\bar{l}_{L_e})_1H_L(YS)_1+\beta_{LS}(\bar{l}_{L_\mu})_{1^{''}}H_L(YS)_{1^{'}}+\gamma_{LS}(\bar{l}_{L_\tau})_{1^{'}}H_L(YS)_{1^{''}}
\end{equation}
\par Now, the mass matrix is given by
\begin{equation}
	m_{LS}=v_{L}
	\begin{pmatrix}
		\alpha_{LS} & 0 & 0\\
		0 & \beta_{LS} & 0\\
		0 & 0 & \gamma_{LS}\\
	\end{pmatrix}
\begin{pmatrix}
	Y_1 & Y_3 & Y_2\\
	Y_2 & Y_1 & Y_3\\
	Y_3 & Y_2 & Y_1\\
\end{pmatrix}
\end{equation}
where $(\alpha_{LS},\beta_{LS},\gamma_{LS})$ are the free parameters and $v_{L}$ is the vev for the left-handed Higgs doublet.
\subsection{Mixing between $\nu_R$ and $S$}
\par In this mixing, the right-handed lepton $l_R$ and sterile fermion $S$ are transformed as triplet in $A_4$ modular group. The modular weight assigned to them is $-1$. The scalar field $H_R$ transformed as singlet under this group.
\par The Lagrangian for this mixing is given by
\begin{equation}
	\mathcal{L}=\alpha_{RS}YH_R(\bar{l}_RS)_{\text{sym}}+\beta_{RS}YH_R(\bar{l}_RS)_{\text{anti-sym}}
\end{equation}
\par In the above equation, the first and the second term represent the symmetric and anti-symmetric part for the product of $l_RS$, making it a triplet under the $A_4$ modular group.
\par The resultant mass matrix is found to be
\begin{equation}
	m_{RS}=v_{R}
	\Biggl(
	\frac{\alpha_{RS}}{3}
	\begin{pmatrix}
		2Y_1 & -Y_3 & -Y_2\\
		-Y_3 & 2Y_2 & -Y_1\\
		-Y_2 & -Y_1 & 2Y_3\\
	\end{pmatrix}
+\frac{\beta_{RS}}{2}
\begin{pmatrix}
	0 & Y_3 & -Y_2\\
	-Y_3 & 0 & Y_1\\
	Y_2 & -Y_1 & 0\\
\end{pmatrix}
\Biggr)
\end{equation}
where $(\alpha_{RS},\beta_{RS})$ are the free parameters and $v_{R}$ is the vev for right-handed Higgs doublet. It is important to observe that $\frac{\alpha_{RS}}{3}$$\neq\frac{\beta_{RS}}{2}$, otherwise the matrix $m_{RS}$ becomes singular, which spoils the purpose of implementing linear seesaw framework.
\subsection{Non-unitarity}
\par Here, we briefly examine the non-unitarity of the neutrino mixing matrix $U_{PMNS}^{'}$. The standard parametrization for $U_{PMNS}^{'}$ takes the form \cite{forero2011lepton}
\begin{equation}
	U_{PMNS}^{'}=\biggl(1-\frac{1}{2}FF^{\dagger}\biggr)U_{PMNS}
\end{equation}
where $U_{PMNS}$ is the PMNS mixing matrix that diagonalises the light neutrino mass matrix. 
\par In this case, the hermitian matrix $F$ is shown to be approximately $F\equiv m_{RS}^{-1}m_{LR}\approx \frac{v_L}{v_R}$. The mass of the $W$ boson $M_W$, the Weinberg angle $\theta_W$, various ratios of the fermionic decays of the $Z$ boson and its invisible decay, electroweak universality, CKM unitarity bounds, lepton flavor violations\cite{fernandez2016global}, and other experimental results are used to determine the global constraints on non-unitarity parameters\cite{antusch2014non,blennow2017non}. The outcome of the experiment is provided by\cite{agostinho2018can}.
\begin{equation}
	|FF^\dagger|\leq
	\begin{bmatrix}
		4.08\times10^{-5} & 1.65\times10^{-5} & 5.19\times10^{-5} \\
		1.65\times10^{-5} & 3.85\times10^{-5} & 5.04\times10^{-5} \\
		5.19\times10^{-5} & 5.04\times10^{-5} & 1.12\times10^{-4} \\
	\end{bmatrix}
\end{equation}
\par In our model, $(\frac{v_L}{v_R})^2\approx10^{-4}$ and $|FF^\dagger|\leq10^{-4}$, which is completely safe for the given non-unitarity bounds stated above.
\section{Numerical Analysis} \label{sec_5}
\textbf{Neutrino mass and mixing}
\par Here, we will perform the numerical analysis for parameters that satisfy the current data on neutrino oscillations~\cite{gonzalez2021nufit}. Table \ref{tab_4} provides the neutrino oscillation data for the $3\sigma$ range.
\begin{table}[H]
\begin{center}
	\caption{Current neutrino oscillation parameters from global fits\cite{gonzalez2021nufit}}
	\begin{tabular}{|c|c|c|c|c|}
		\hline
				Oscillation parameters &  \multicolumn{2}{c|}{Normal Ordering}  &\multicolumn{2}{c|}{Inverted Ordering} \\
		\cline{2-5}
	         	&  Best fit$\pm 1\sigma$       & $3\sigma$ range   & Best fit$\pm1\sigma$   & $3\sigma$ range       \\
		\hline
		$\Delta m_{21}^2/10^{-5}{eV^2}$ & $7.42_{-0.20}^{+0.21}$ & $6.82-8.04$ & $7.42_{-0.20}^{+0.21}$ & $6.82-8.04$\\
		\hline
		$|\Delta m_{3l}^2|/10^{-3}{eV^2}$ & $2.510_{-0.027}^{+0.027}$ &  $2.430-2.593$   & $-2.490_{-0.028}^{+0.026}$  & $-2.574--2.410$ \\
		\hline
		$\theta_{12}^\circ$ & $33.45_{-0.75}^{+0.77}$ &  $31.27-35.87$   &  $33.45_{-0.75}^{+0.78}$   & $31.27-35.87$ \\
		\hline
		$\theta_{23}^\circ$ & $42.1_{-0.9}^{+1.1}$ & $39.7-50.9$  & $49.0_{-1.9}^{+0.9}$   & $39.8-51.6$ \\
		\hline
		$\theta_{13}^\circ$ & $8.62_{-0.12}^{+0.12}$ &  $8.25-8.98$   &  $8.61_{-0.12}^{+0.14}$   & $8.24-9.02$ \\
		\hline
		$\delta_{CP}^\circ$ & $230_{-25}^{+36}$ & $144-350$ & $278_{-30}^{+22}$ & $194-345$\\
		\hline
	\end{tabular}
 \label{tab_4}
\end{center}
\end{table}

\par Here, the neutrino mass matrix of Eqn. \eqref{eqn_9} can be numerically diagonalized by a unitary matrix U, from which the standard relations can be used to obtain the neutrino mixing angles:
\begin{equation}
	\sin^2\theta_{13}=|U_{13}|^2, \; \sin^2\theta_{12}=\frac{|U_{12}^2|}{1-|U_{13}^2|}, \; \sin^2\theta_{23}=\frac{|U_{23}^2|}{1-|U_{13}^2|} \label{eqn_33}
\end{equation}
\par We can choose the following ranges for the free parameters of the model to satisfy the current neutrino oscillation data:
\begin{align*}
	\text{Re[$\tau$]}\in[0.3,1.8], \; \text{Im[$\tau$]}\in[0.75,2.5], \; \{\alpha_{LS},\beta_{LS},\gamma_{LS}\}\in10^{-3}[0.1,1], \\
	\alpha_{RS}\in[1,5], \; \beta_{RS}\in[0,0.001], \; v_{LS}\in[10^3,10^4]\text{ eV}, \; v_{RS}\in[1,10]\text{ TeV}, \Lambda\in[10,100]\text{ TeV}.
\end{align*}
\par
\begin{figure}[H]
	\centering
	\begin{tabular}{cccc}
		\includegraphics[width=0.45\textwidth]{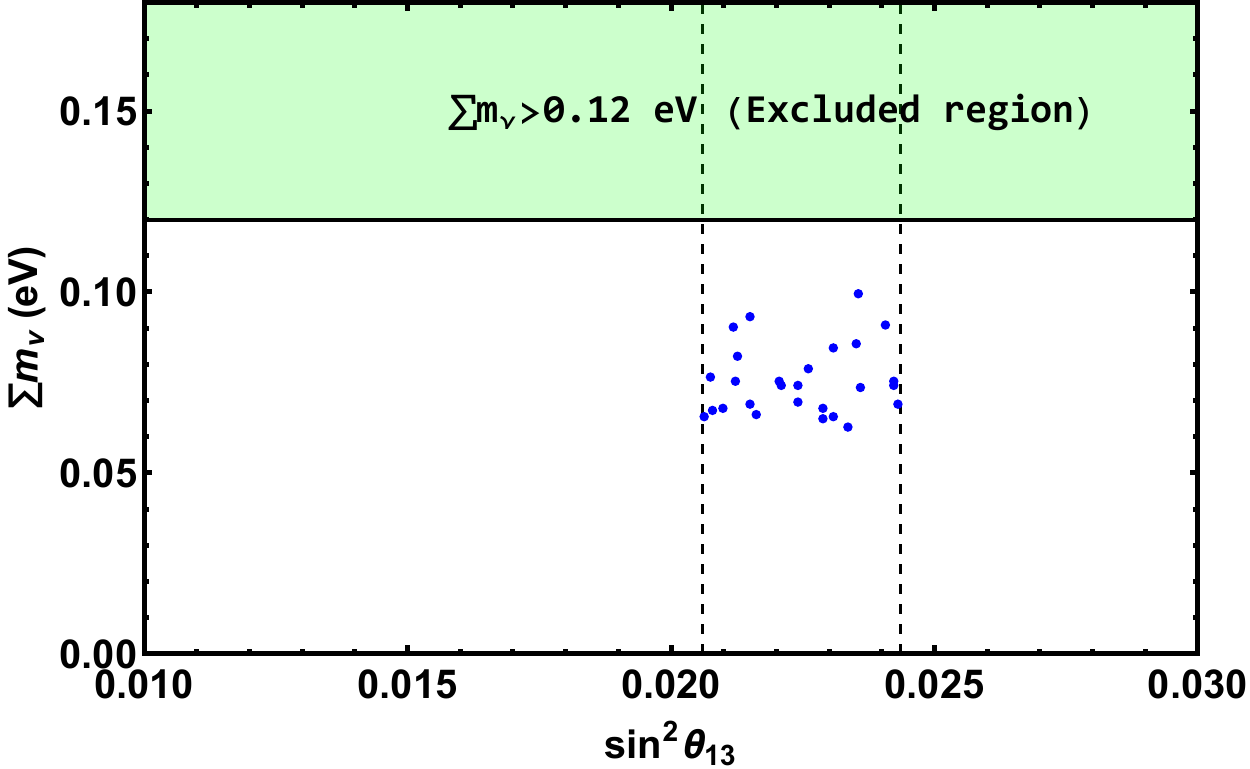} &
		\includegraphics[width=0.45\textwidth]{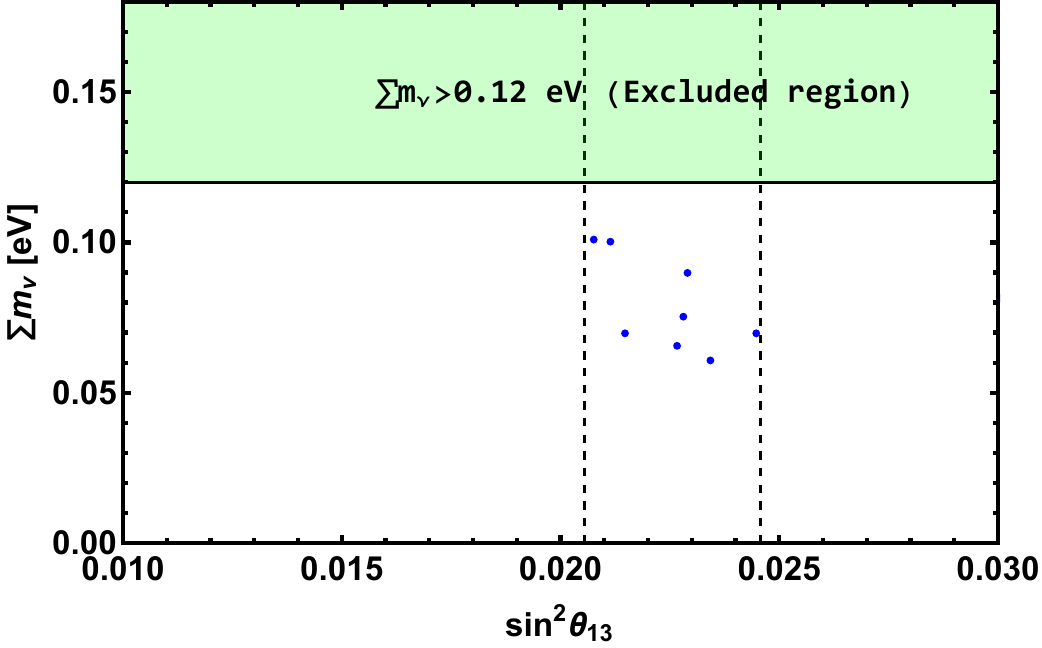} \\
	\end{tabular}
	\caption{The correlation plot for neutrino masses $(\Sigma m_\nu)$ with $\sin^2\theta_{13}$ for NO (left) and IO (right).}
	\label{fig:plot1}
\end{figure}

\begin{figure}[H]
	\centering
	\begin{tabular}{cccc}
		\includegraphics[width=0.45\textwidth]{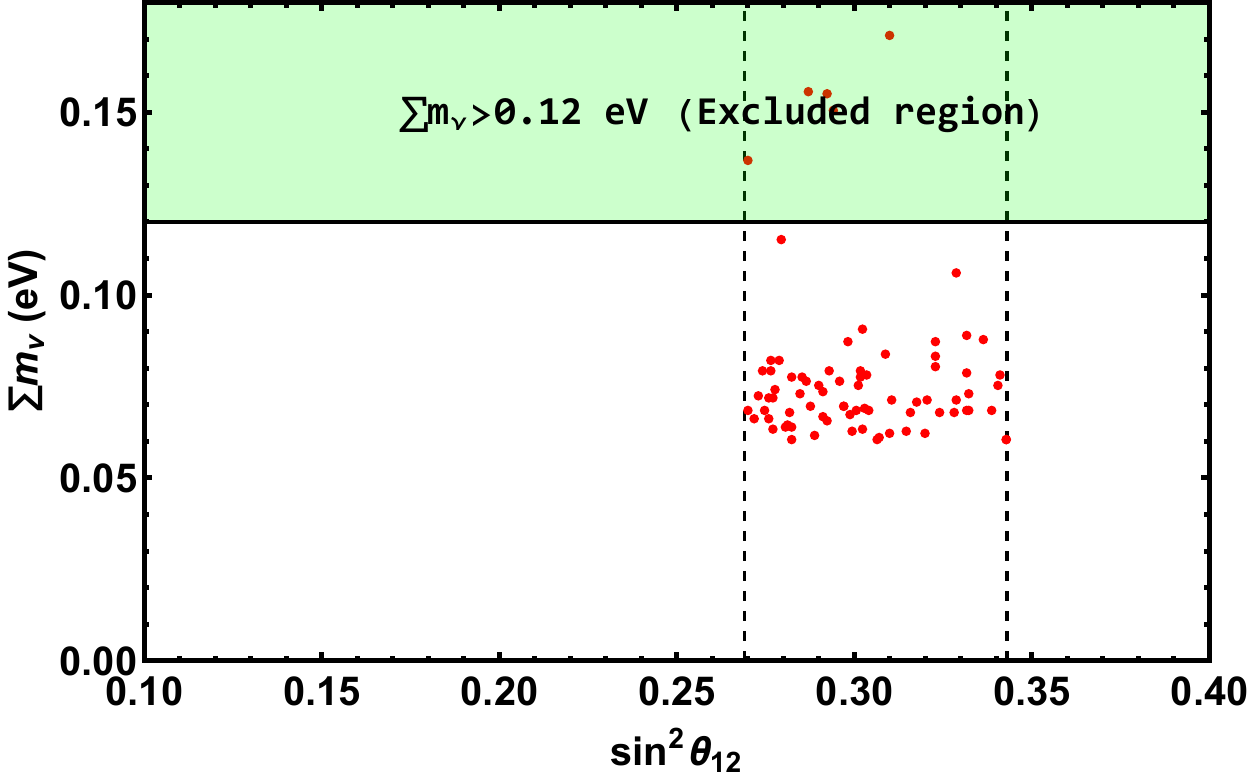} &
		\includegraphics[width=0.45\textwidth]{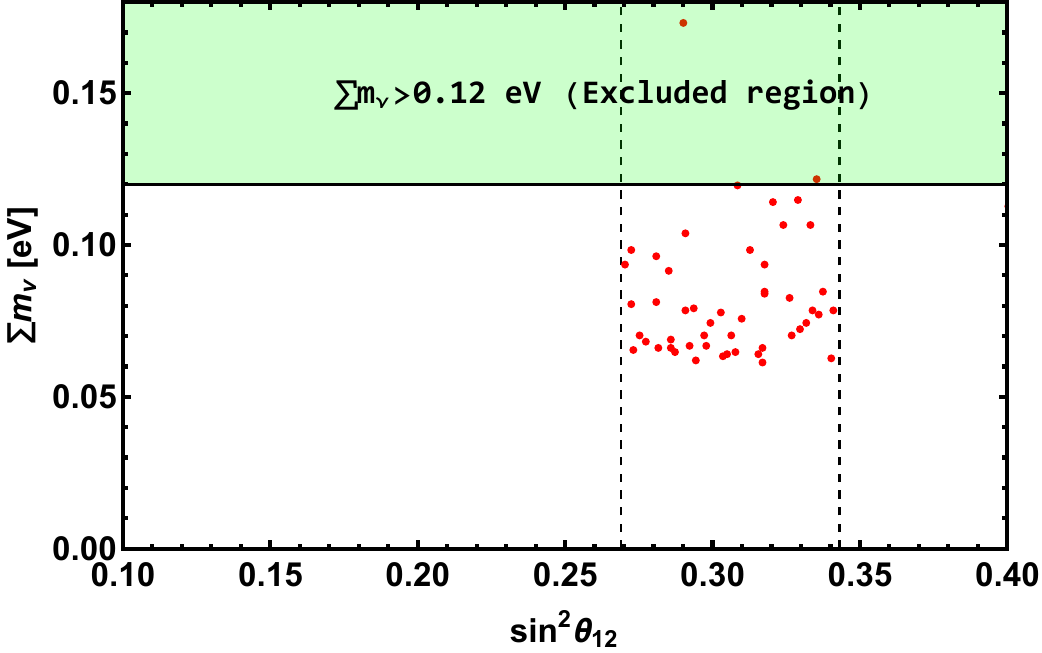} \\
	\end{tabular}
	\caption{The correlation plot for sum of neutrino masses $(\Sigma m_\nu)$ with $\sin^2\theta_{12}$ for NO (left) and IO (right).}
	\label{fig:plot2}
\end{figure}

\begin{figure}[H]
	\centering
	\begin{tabular}{cccc}
		\includegraphics[width=0.45\textwidth]{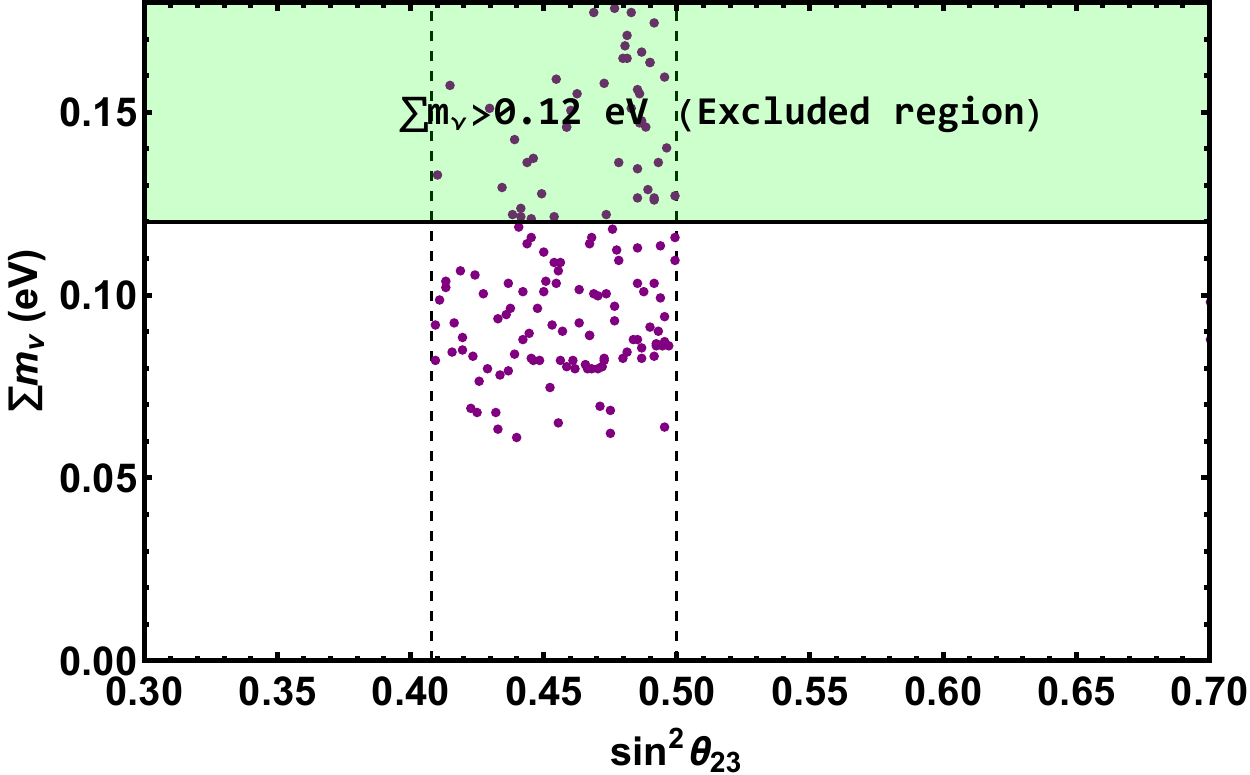} &
		\includegraphics[width=0.45\textwidth]{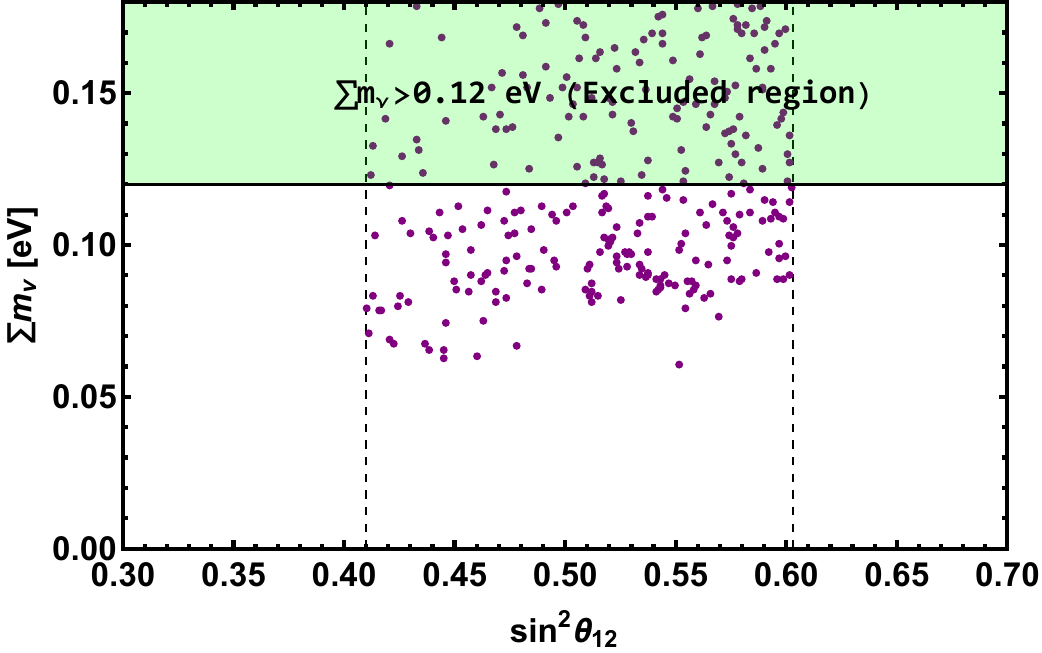} \\
	\end{tabular}
	\caption{The correlation plot for sum of neutrino masses $(\Sigma m_\nu)$ with $\sin^2\theta_{23}$ for NO (left) and IO (right).}
	\label{fig:plot3}
\end{figure}

 The relations provided in Eqn. \eqref{eqn_33} can be used to get the mixing angles for the model. The variations of the sum of neutrino masses $(\Sigma m_\nu)$ with the mixing angles $\sin^2\theta_{13}$, $\sin^2\theta_{12}$, and $\sin^2\theta_{23}$ for both normal and inverted ordering are shown in figures \ref{fig:plot1}, \ref{fig:plot2}, and \ref{fig:plot3}, respectively.

\par The figures show that a wide range of parameter spaces are within the permitted range for the upper bound of neutrino masses $(\Sigma m_\nu=0.12 \text{ eV})$ and the mixing angles. For both NO and IO, the lower bound for the sum of neutrino masses is found to be approximately $0.06 \text{ eV}$.
\par
\begin{figure}[H]
	\centering
	\begin{tabular}{cccc}
		\includegraphics[width=0.45\textwidth]{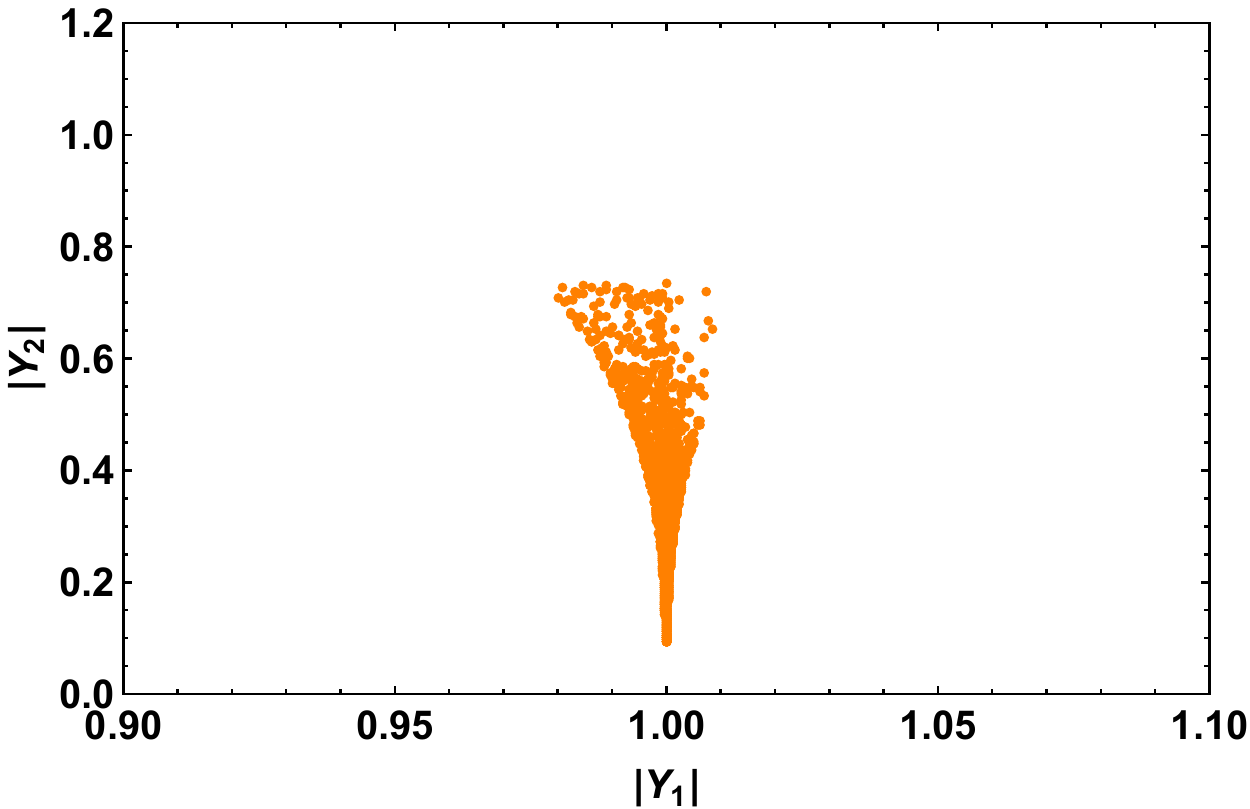} &
		\includegraphics[width=0.45\textwidth]{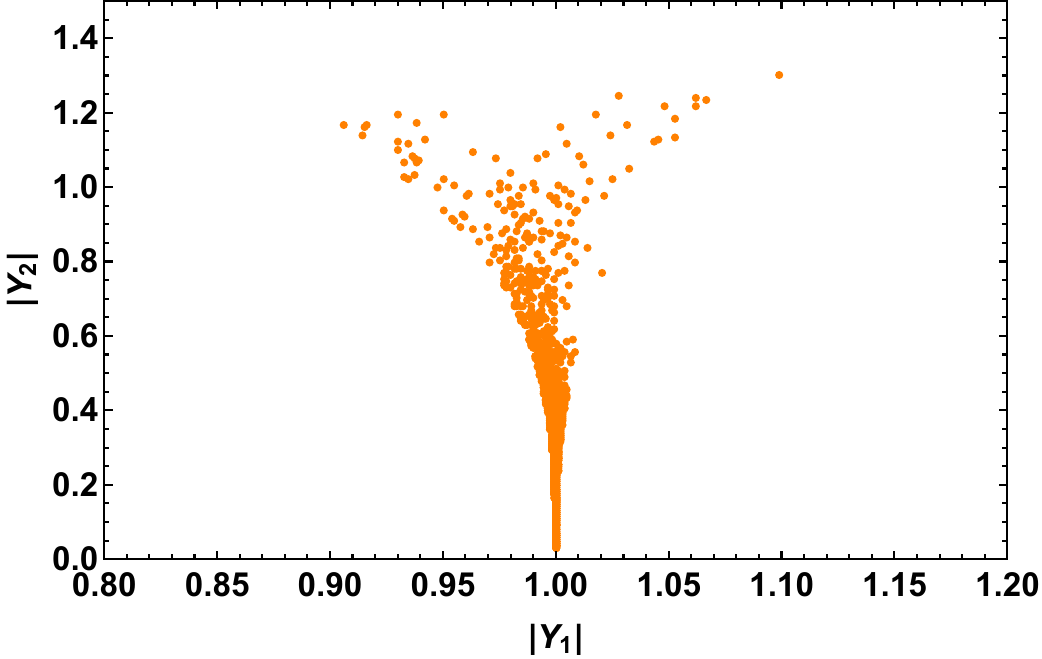} \\
	\end{tabular}
	\caption{Variation of Yukawa couplings $|Y_1|$ and $|Y_2|$ for NO (left) and IO (right).}
	\label{fig:plot4}
\end{figure}

\begin{figure}[H]
	\centering
	\begin{tabular}{cccc}
		\includegraphics[width=0.45\textwidth]{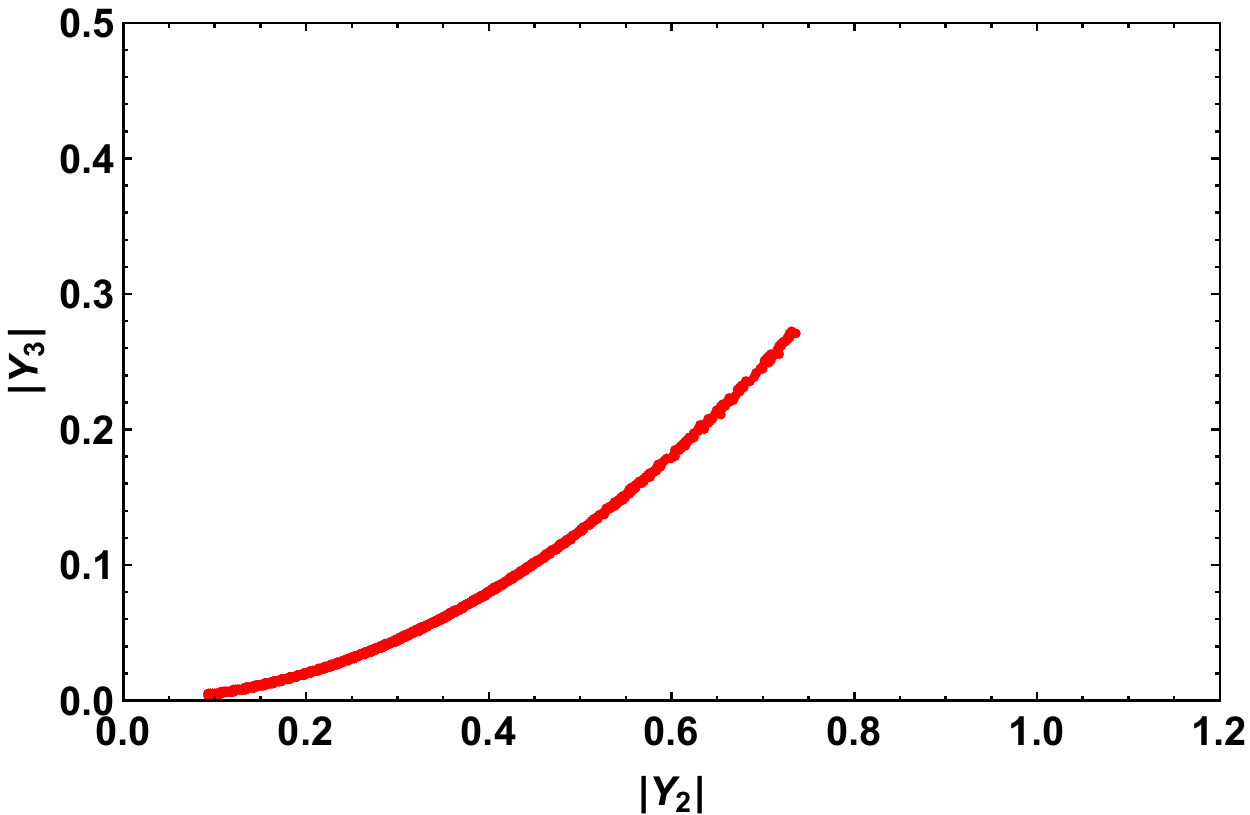} &
		\includegraphics[width=0.45\textwidth]{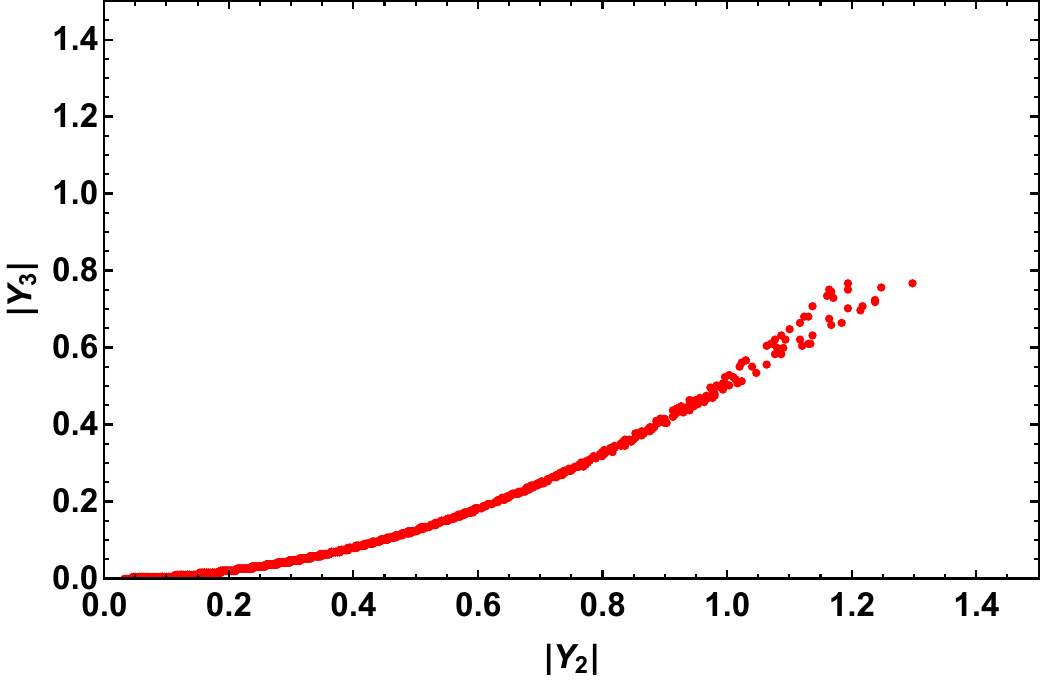} \\
	\end{tabular}
	\caption{Variation of Yukawa couplings $|Y_2|$ and $|Y_3|$ for NO (left) and IO (right).}
	\label{fig:plot5}
\end{figure}

 The correlation between the magnitudes of Yukawa couplings $Y_1$ and $Y_2$, $Y_2$ and $Y_3$ are shown in figures \ref{fig:plot4} and \ref{fig:plot5} respectively. The ranges for yukawa couplings are given in Table \ref{tab_5}.

\begin{table}[H]
\begin{center}
	\caption{Ranges for Yukawa couplings obtained from the model.}
	\begin{tabular}{|c|c|c|}
		\hline
		Yukawa couplings & Normal Ordering  & Inverted Ordering \\
		\hline
		$|Y_1|$ & $0.98-1.01$ & $0.90-1.05$ \\
		\hline
		$|Y_2|$ & $0.08-0.75$ & $0.04-1.25$ \\
		\hline
		$|Y_3|$ & $0.002-0.26$ & $0.02-0.80$ \\
		\hline
	\end{tabular}
 \label{tab_5}
\end{center}
\end{table}

\par
\begin{figure}[H]
	\centering
	\begin{tabular}{cccc}
		\includegraphics[width=0.45\textwidth]{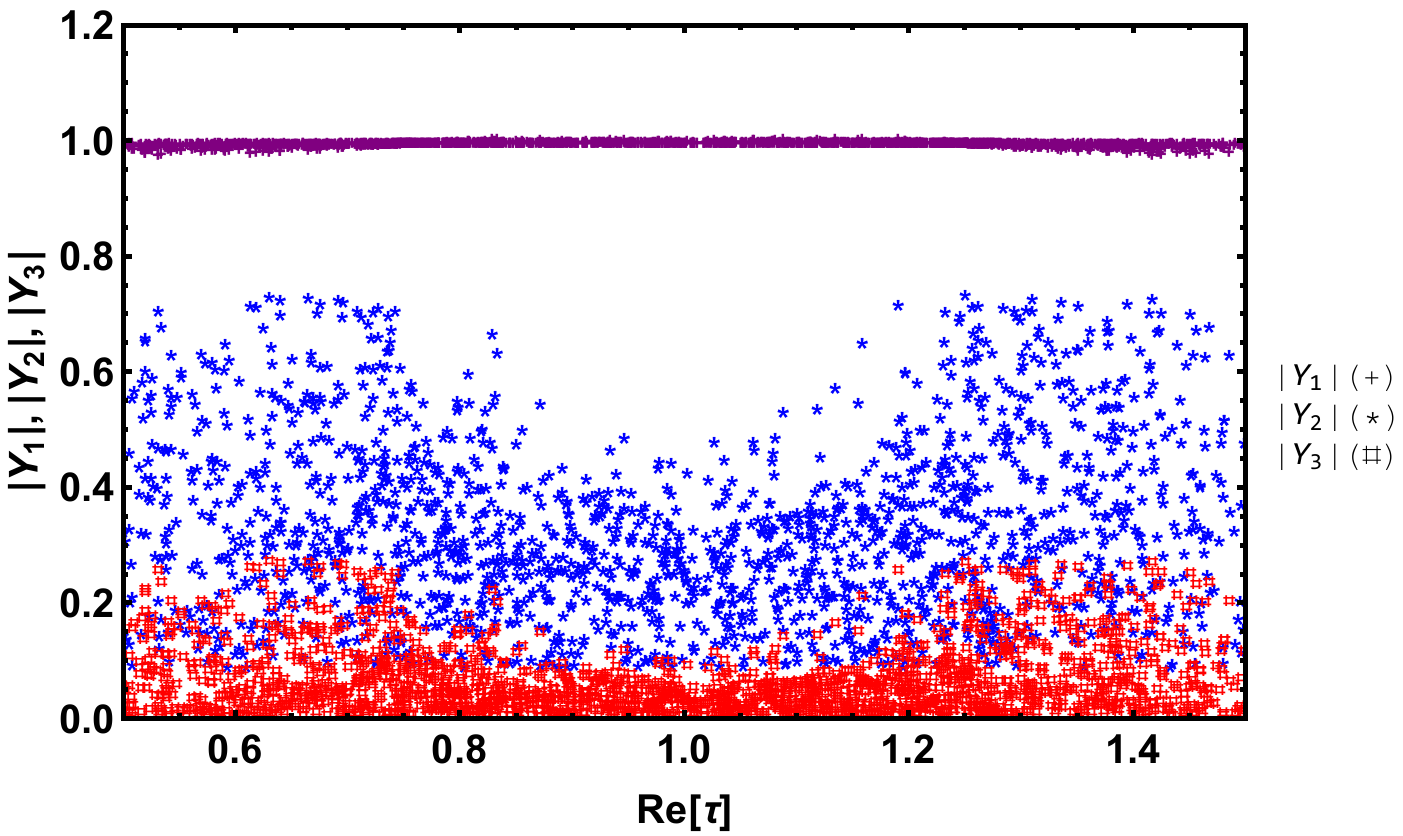} &
		\includegraphics[width=0.45\textwidth]{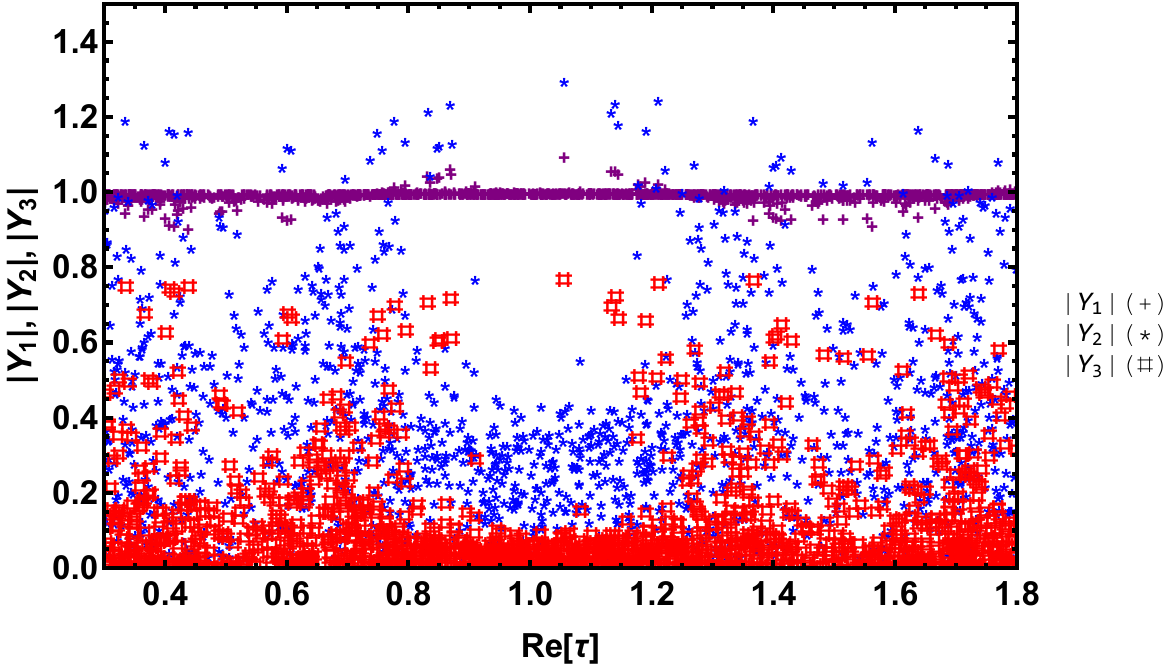} \\
	\end{tabular}
	\caption{Variation of Yukawa couplings as a function of real component of $\tau$ for NO (left) and IO (right).}
	\label{fig:plot6}
\end{figure}

\begin{figure}[H]
	\centering
	\begin{tabular}{cccc}
		\includegraphics[width=0.45\textwidth]{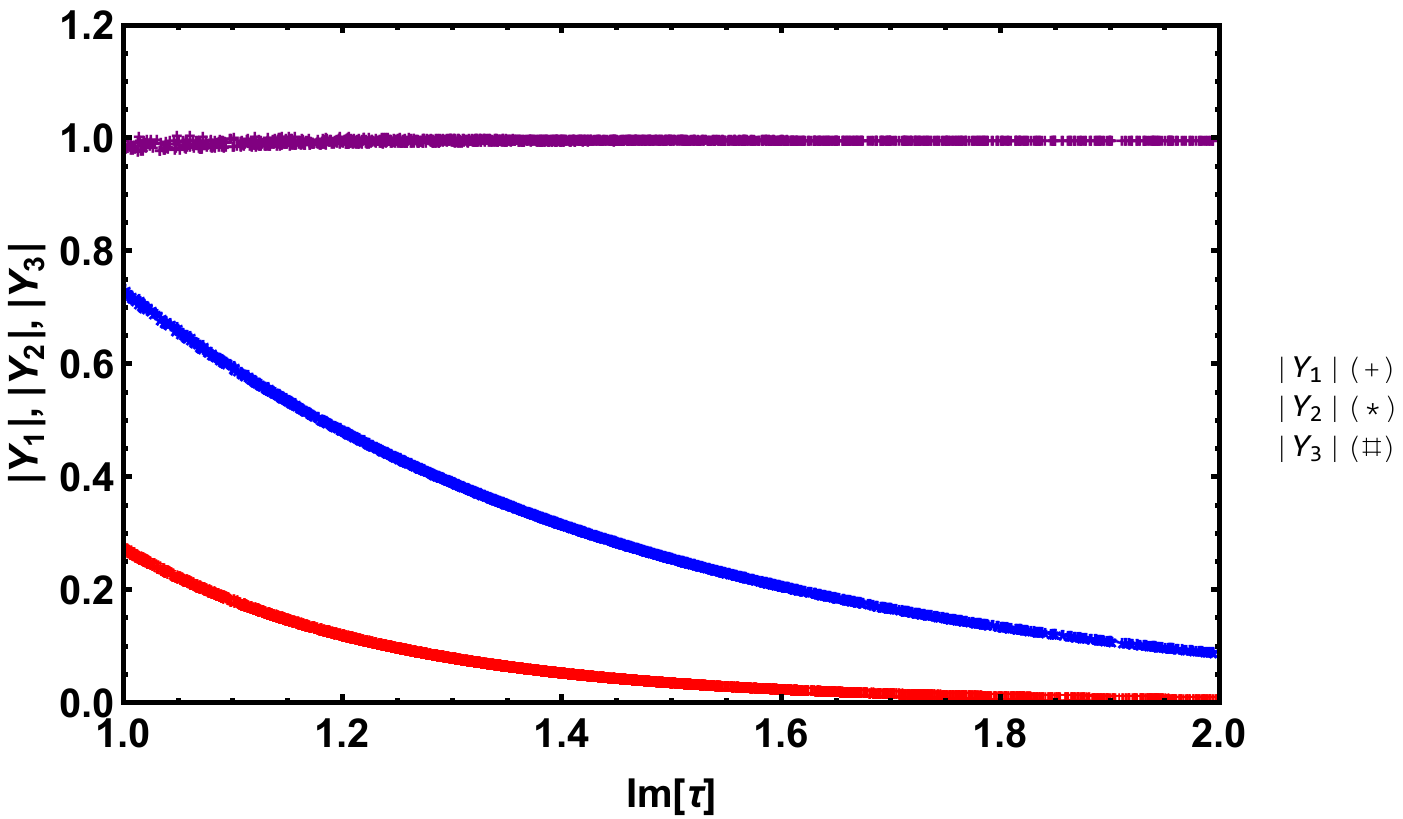} &
		\includegraphics[width=0.45\textwidth]{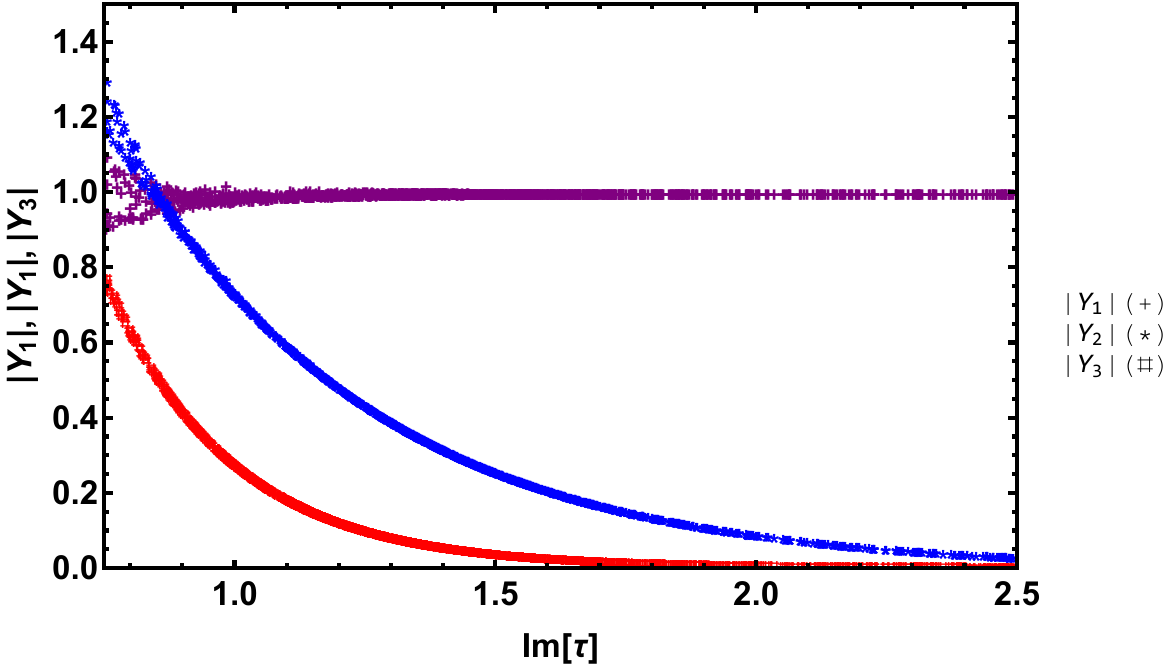} \\
	\end{tabular}
	\caption{Variation between Yukawa couplings as a function of imaginary component of $\tau$ for NO (left) and IO (right).}
	\label{fig:plot7}
\end{figure}

 The variation between Yukawa couplings and the real component of modulus $\tau$ for both NO and IO are shown in figure \ref{fig:plot6}. Furthermore, figure \ref{fig:plot7} shows the variation of Yukawa couplings with the imaginary component of $\tau$.

\par
\begin{figure}[H]
	\centering
	\begin{tabular}{cccc}
		\includegraphics[width=0.45\textwidth]{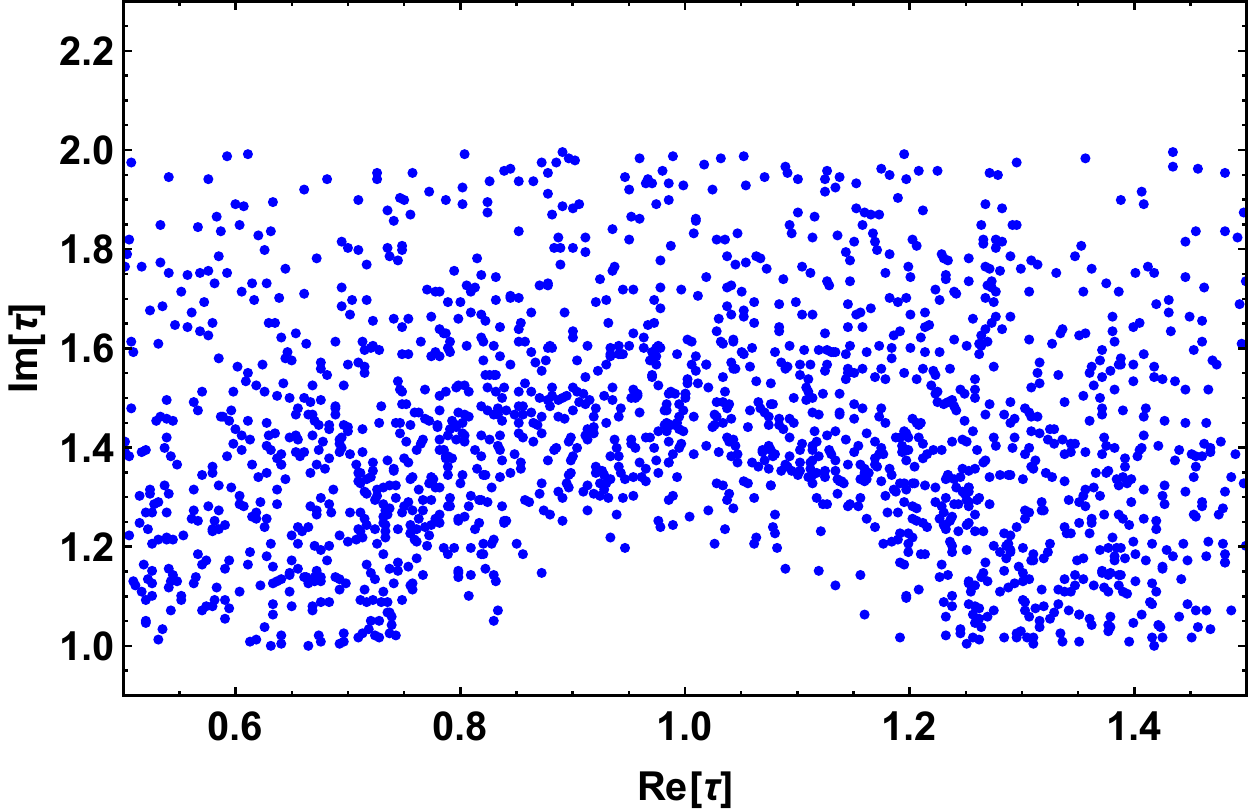} &
		\includegraphics[width=0.45\textwidth]{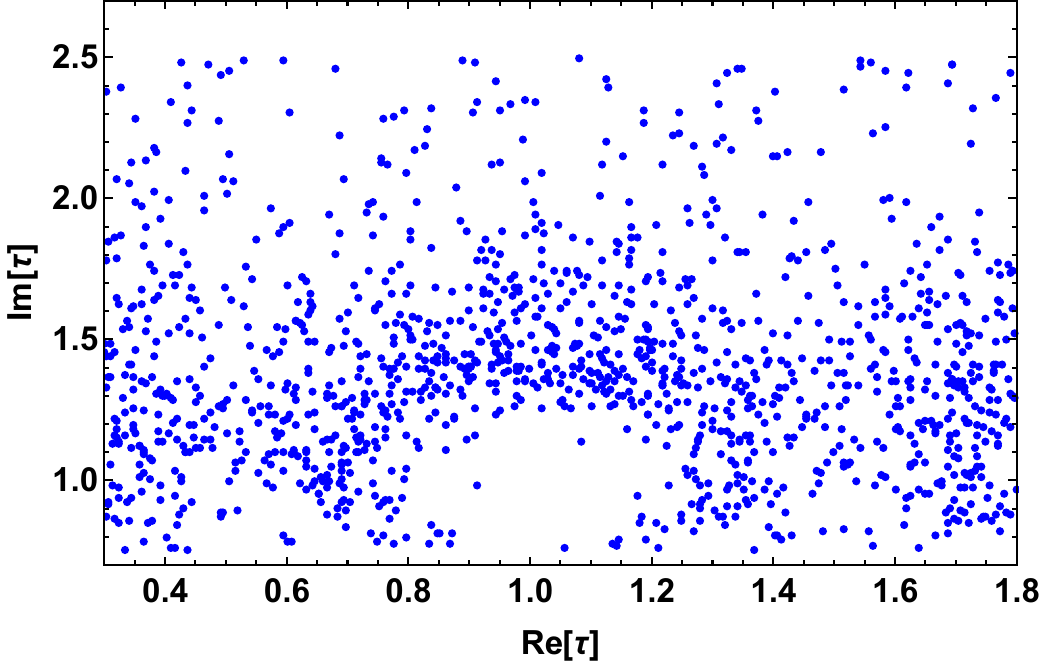} \\
	\end{tabular}
	\caption{The region of modulus $\tau$ satisfying neutrino oscillation data for NO (left) and IO (right).}
	\label{fig:plot8}
\end{figure}

 The region where modulus $\tau$ for both NO and IO is allowed and adheres to all the constraints used to deduce the neutrino oscillation parameters is shown in figure \ref{fig:plot8}.

\par
\begin{figure}[H]
	\centering
	\begin{tabular}{cccc}
		\includegraphics[width=0.45\textwidth]{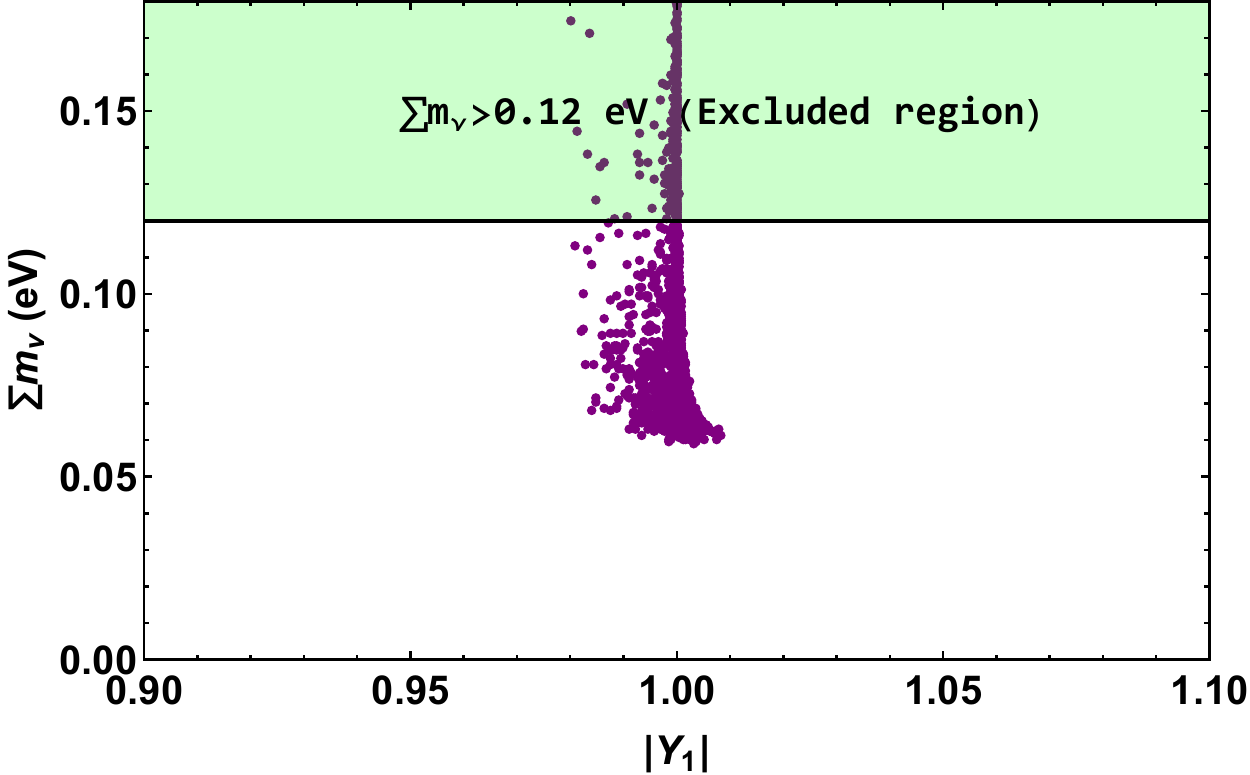} &
		\includegraphics[width=0.45\textwidth]{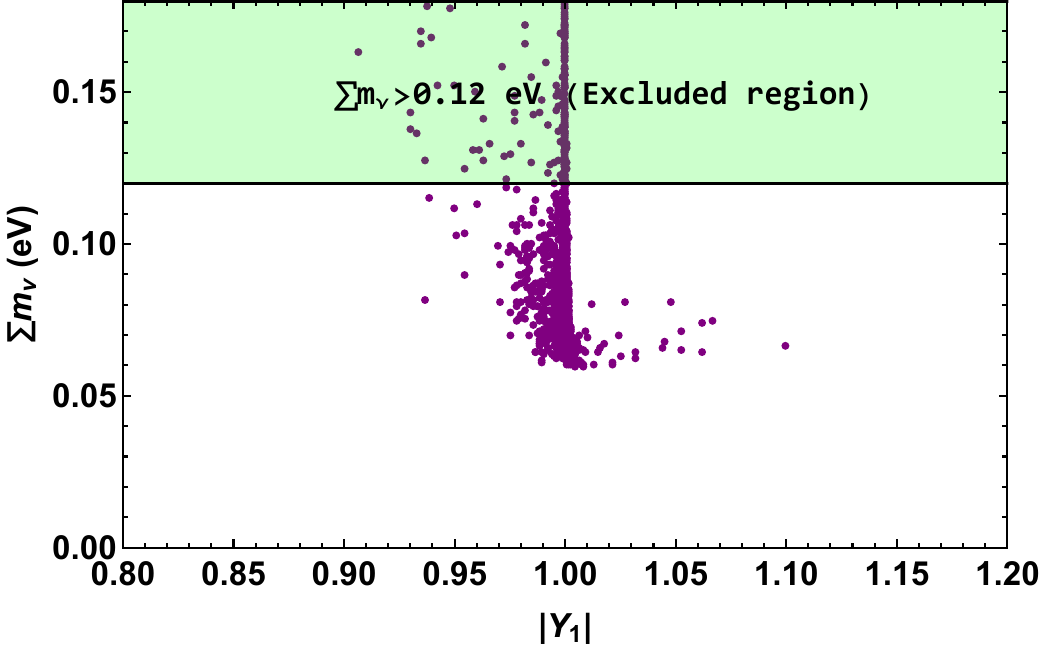} \\
		\includegraphics[width=0.45\textwidth]{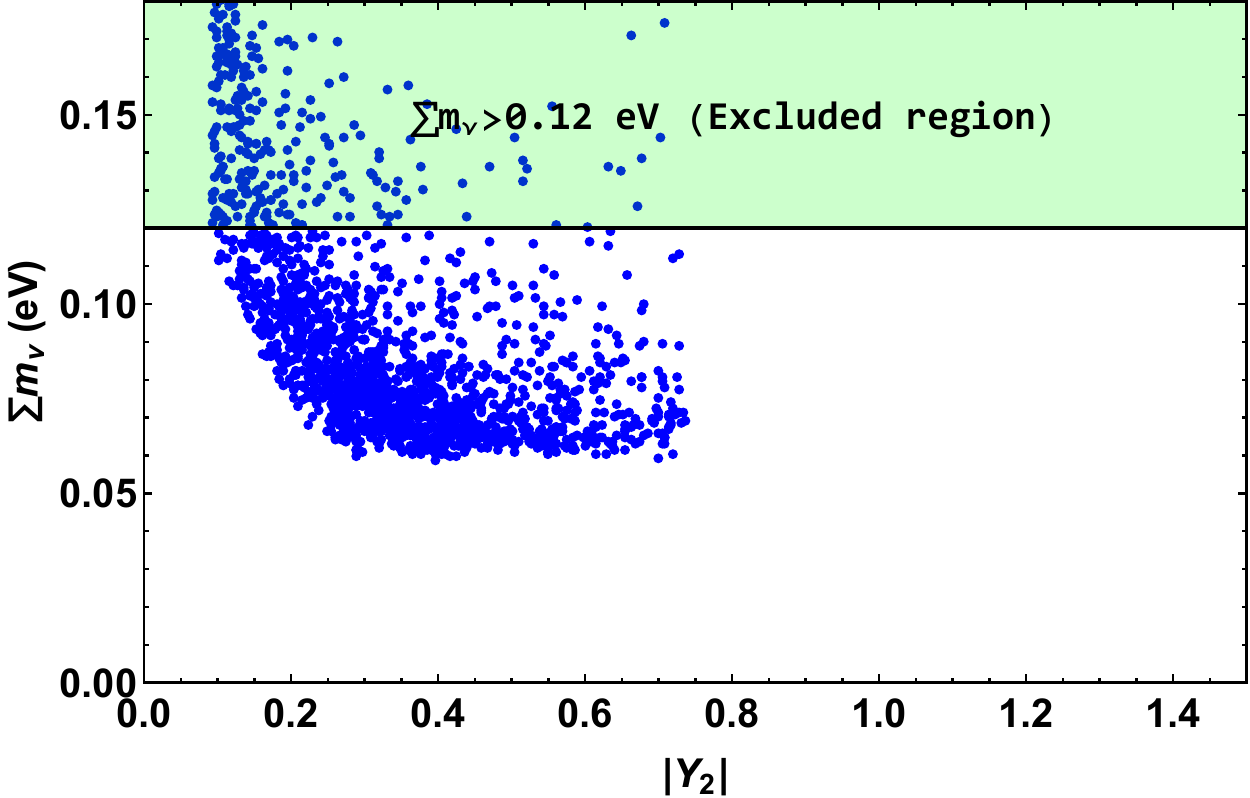} &
		\includegraphics[width=0.45\textwidth]{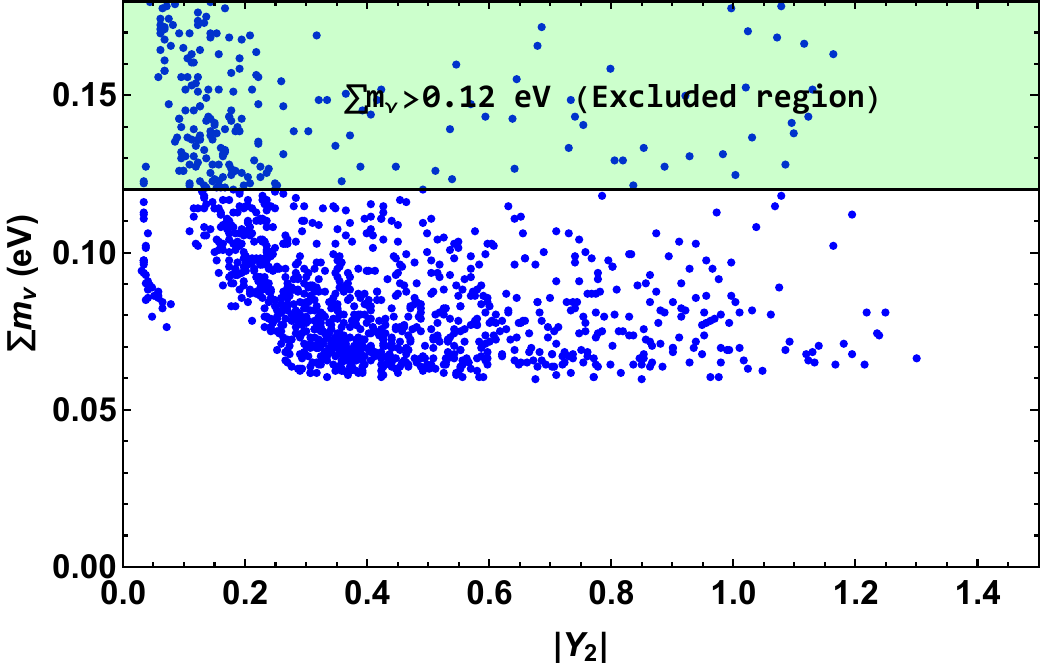} \\
		\includegraphics[width=0.45\textwidth]{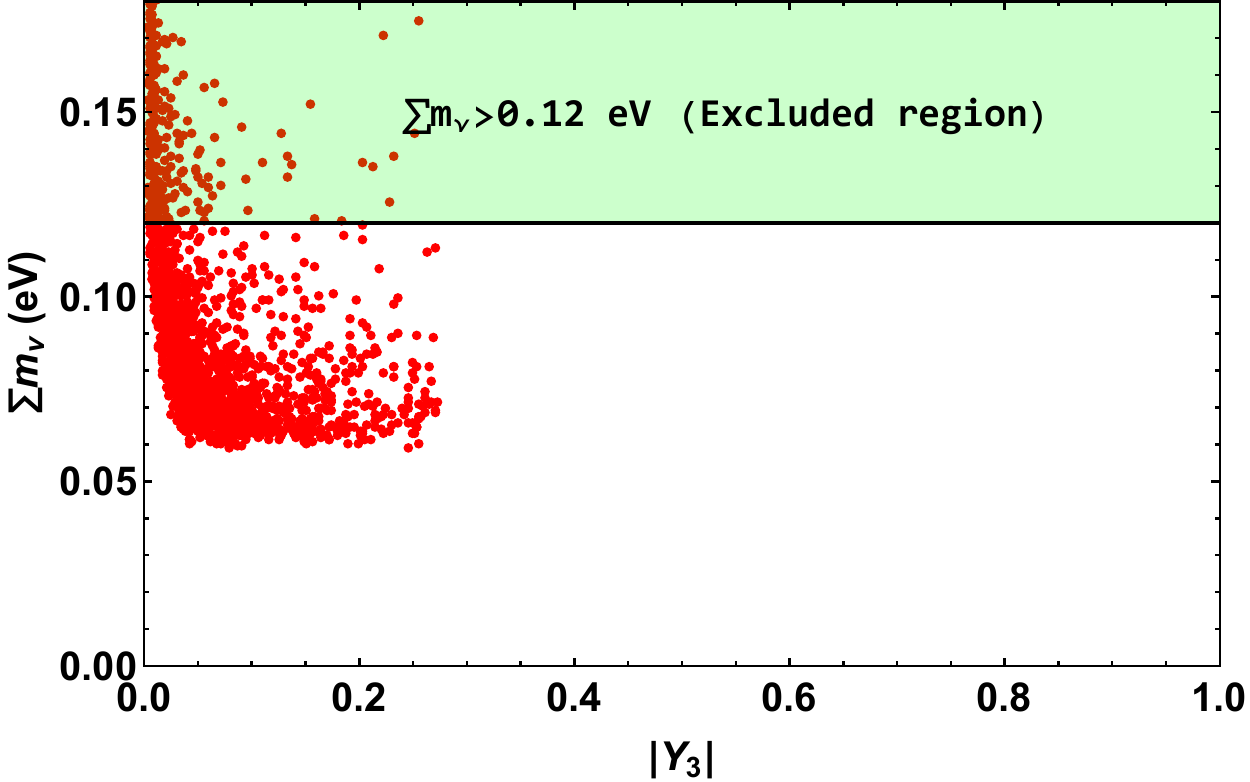} &
		\includegraphics[width=0.45\textwidth]{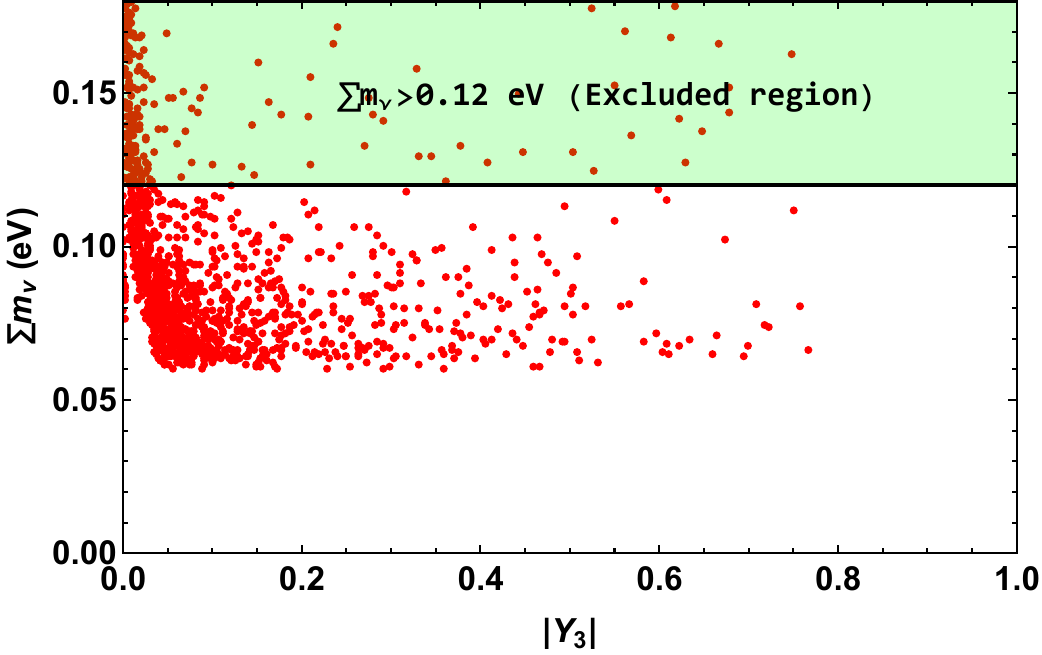} \\
	\end{tabular}
	\caption{Variation of Yukawa coupligs with the sum of active neutrino masses for NO (left) and IO (right).}
	\label{fig:plot9}
\end{figure}

 The parameter space for Yukawa couplings corresponding to the observed sum masses of active neutrinos is shown in figure \ref{fig:plot9}.
\par
\begin{figure}[H]
	\centering
	\begin{tabular}{cccc}
		\includegraphics[width=0.45\textwidth]{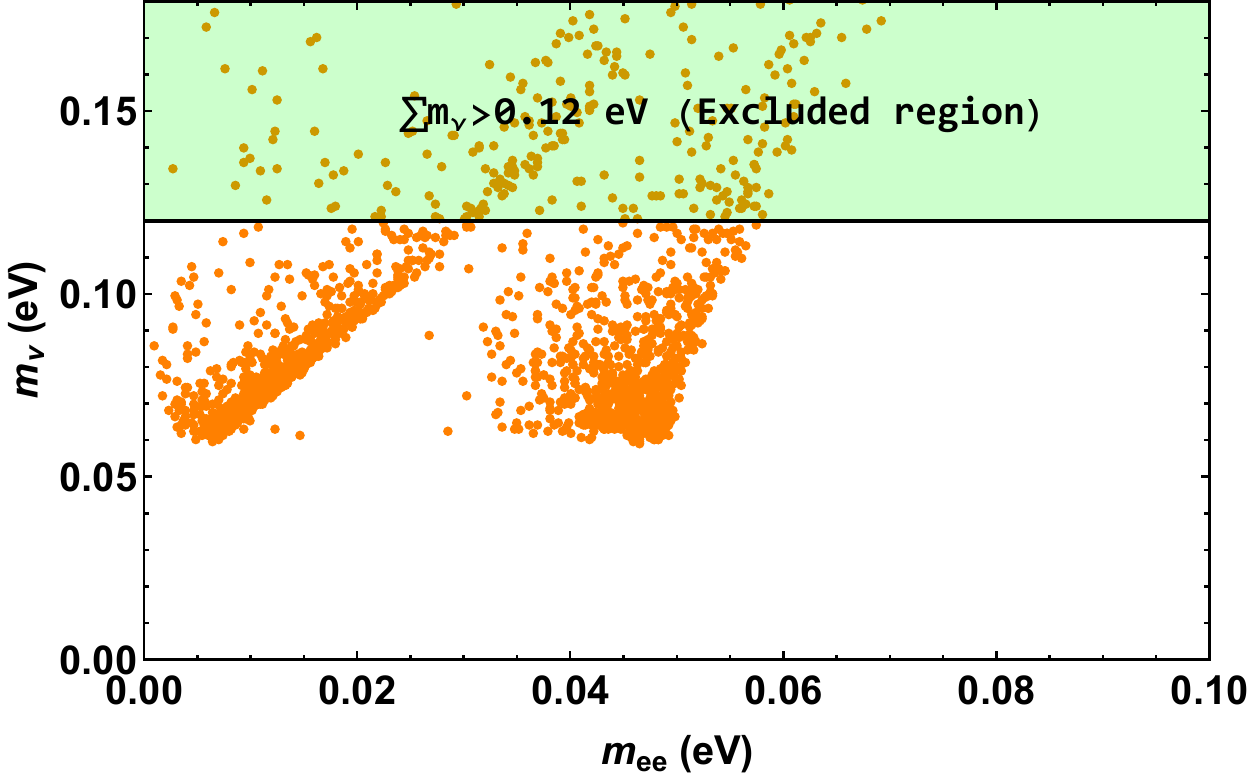} &
		\includegraphics[width=0.45\textwidth]{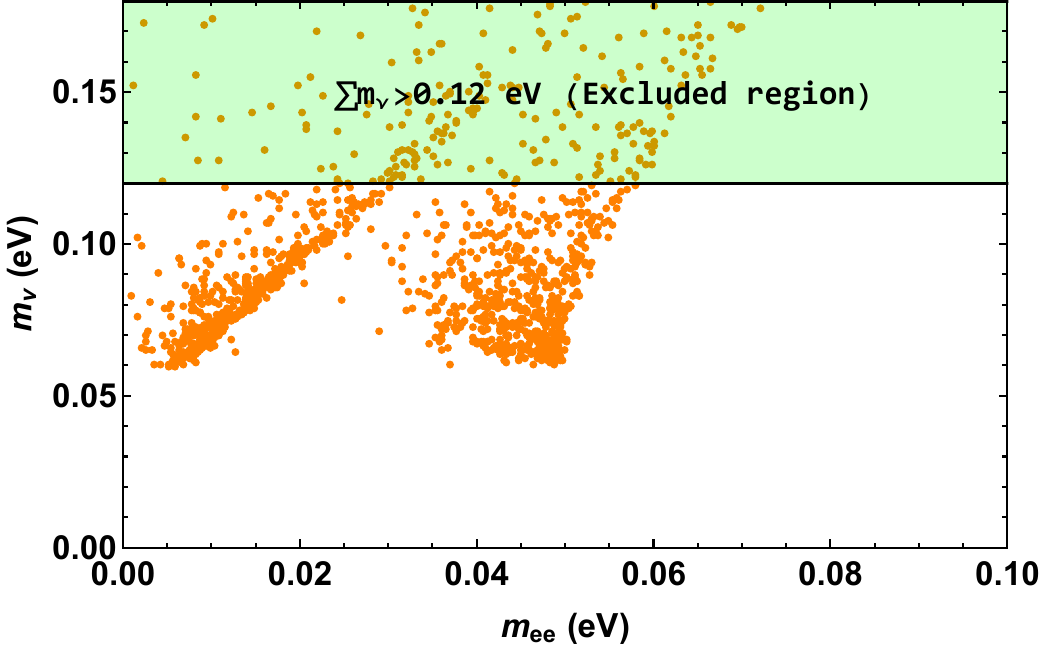} \\
	\end{tabular}
	\caption{Correlation between sum of neutrino masses $(\sum m_\nu)$ versus effective mass for neutrinoless double beta decay $(m_{ee})$ for NO (left) and IO (right).}
	\label{fig:plot10}
\end{figure}

 The correlation between sum of the neutrino masses $(\sum m_\nu)$ versus effective mass for neutrinoless double beta decay $(m_{ee})$ is shown in figure \ref{fig:plot10}.
 \par From the figure, the effective mass $m_{ee}$ is found to be within the experimental bound for both NO and IO.
\begin{figure}[H]
    \centering
    \begin{tabular}{cccc}
    \includegraphics[width=0.45\textwidth]{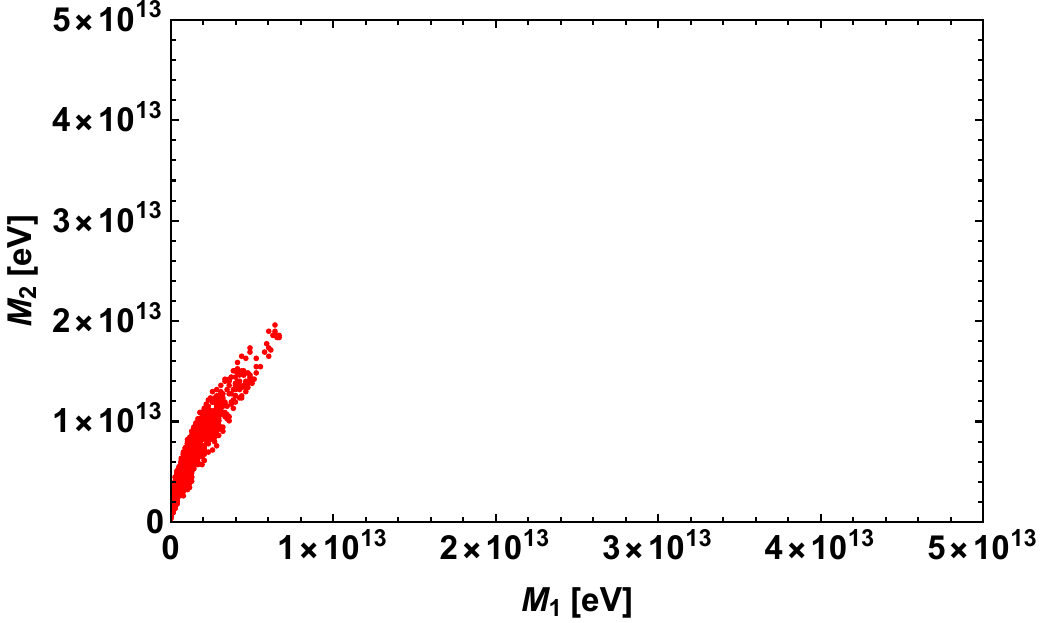} &
    \includegraphics[width=0.45\textwidth]{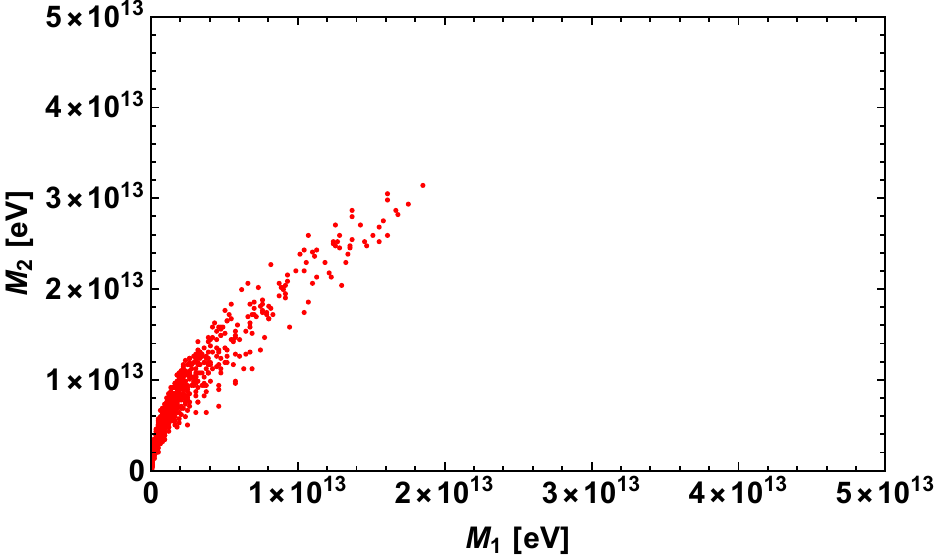} \\
    \includegraphics[width=0.45\textwidth]{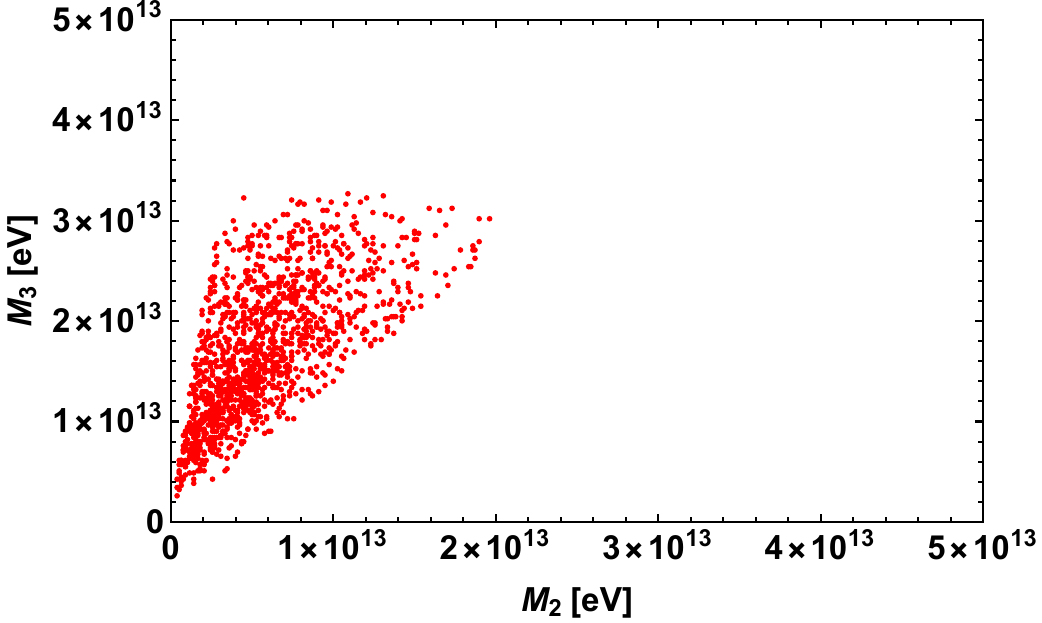} &
    \includegraphics[width=0.45\textwidth]{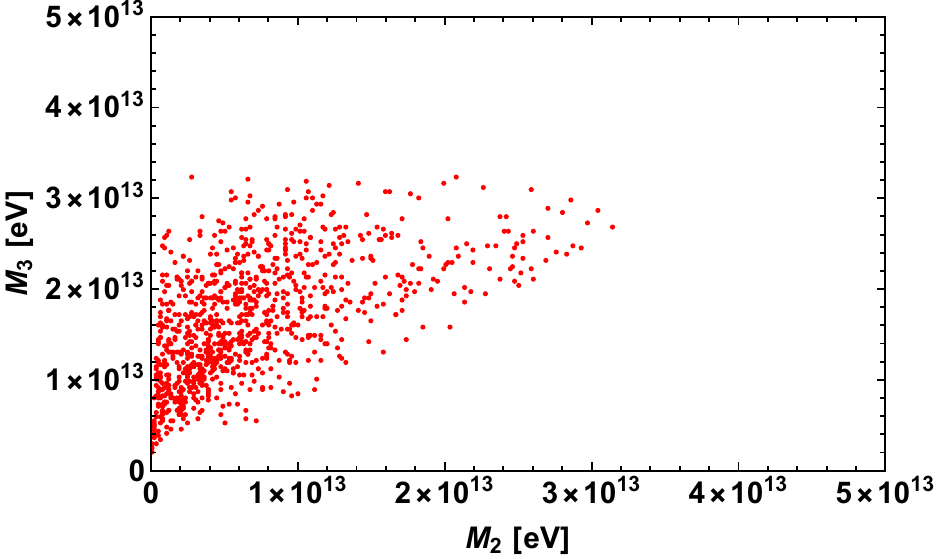} \\
    \end{tabular}
    \caption{Correlation between the masses of heavy neutrinos for NO (left) and IO (right).}
    \label{fig:plot11}
\end{figure}
\par The heavy neutrino masses computed as pseudo Dirac fermions are shown in figure \ref{fig:plot11}. The mass relations are found to be $M_1<<M_2<M_3$ and $M_1<M_2\lesssim M_3$ for NO and IO respectively. Additionally, in case of IO, mass scale tends to be larger.

\begin{figure}[H]
	\centering 
	\begin{tabular}{cccc}
		\includegraphics[width=0.45\textwidth]{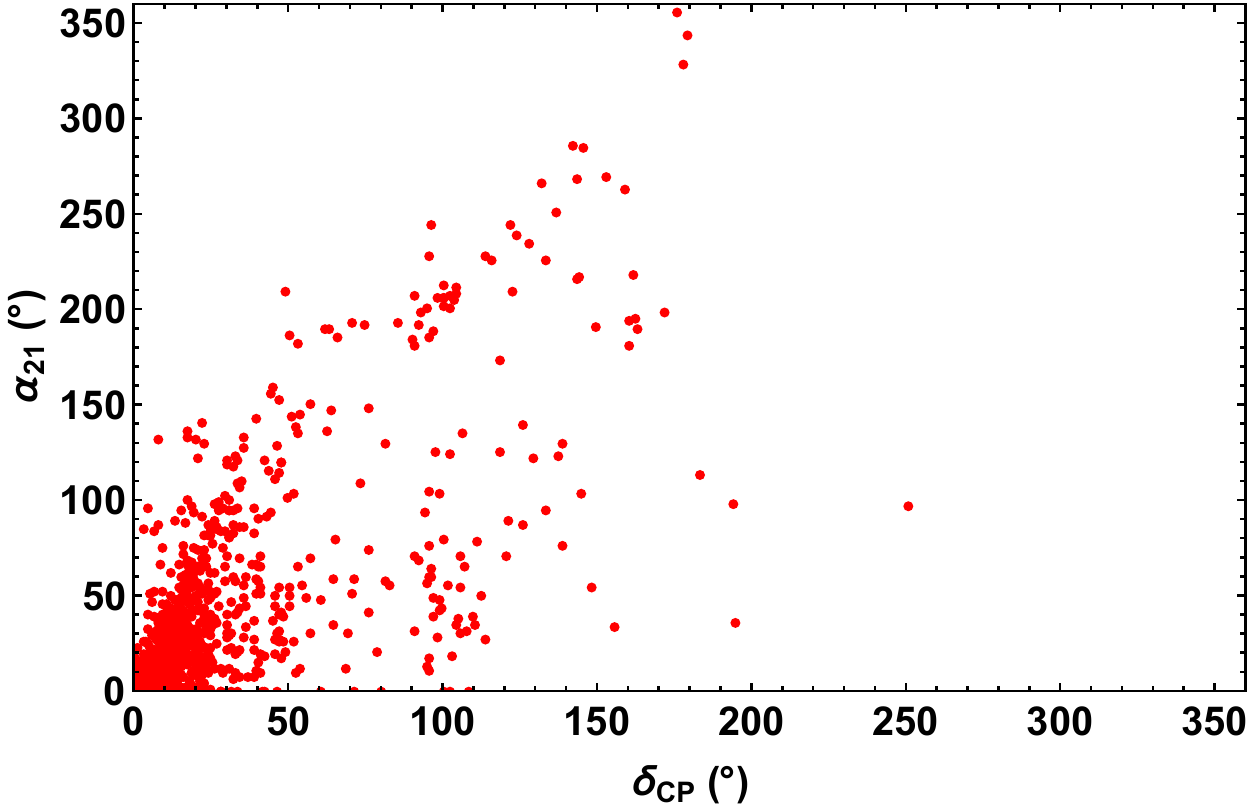} &
		\includegraphics[width=0.45\textwidth]{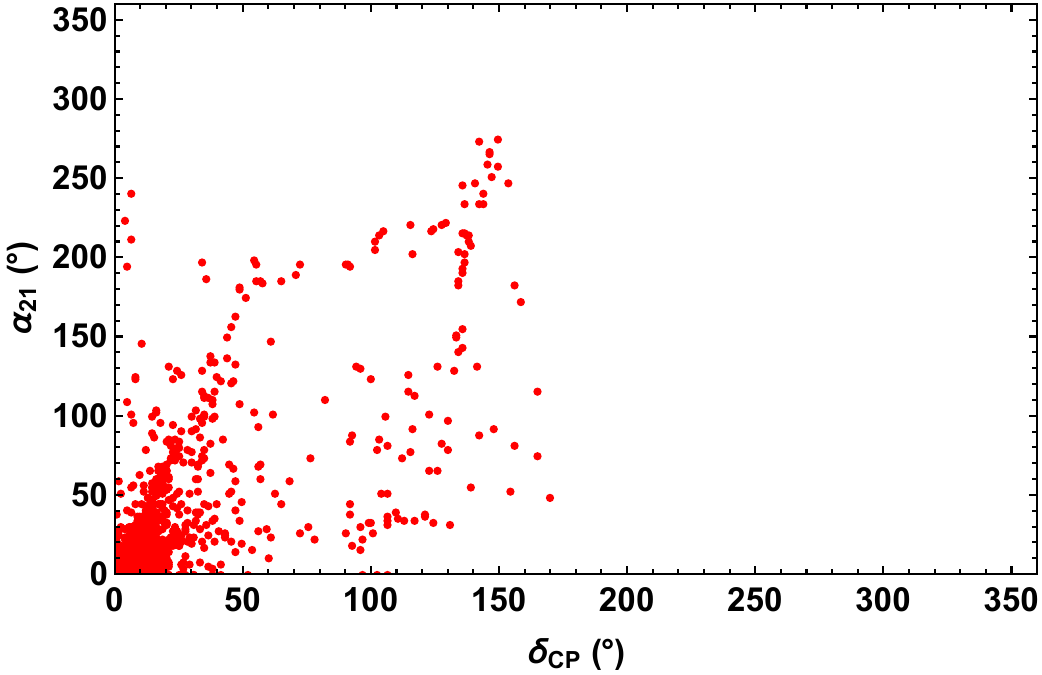} \\
            \includegraphics[width=0.45\textwidth]{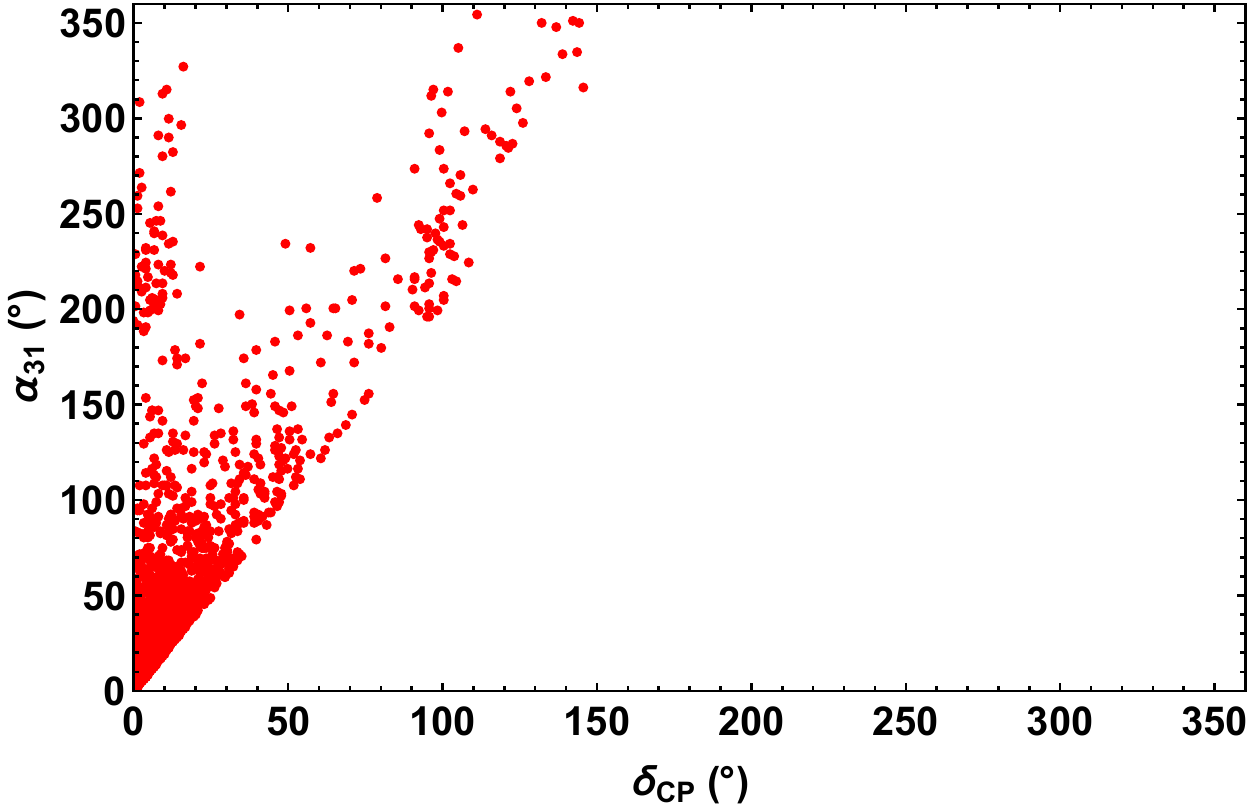} &
		\includegraphics[width=0.45\textwidth]{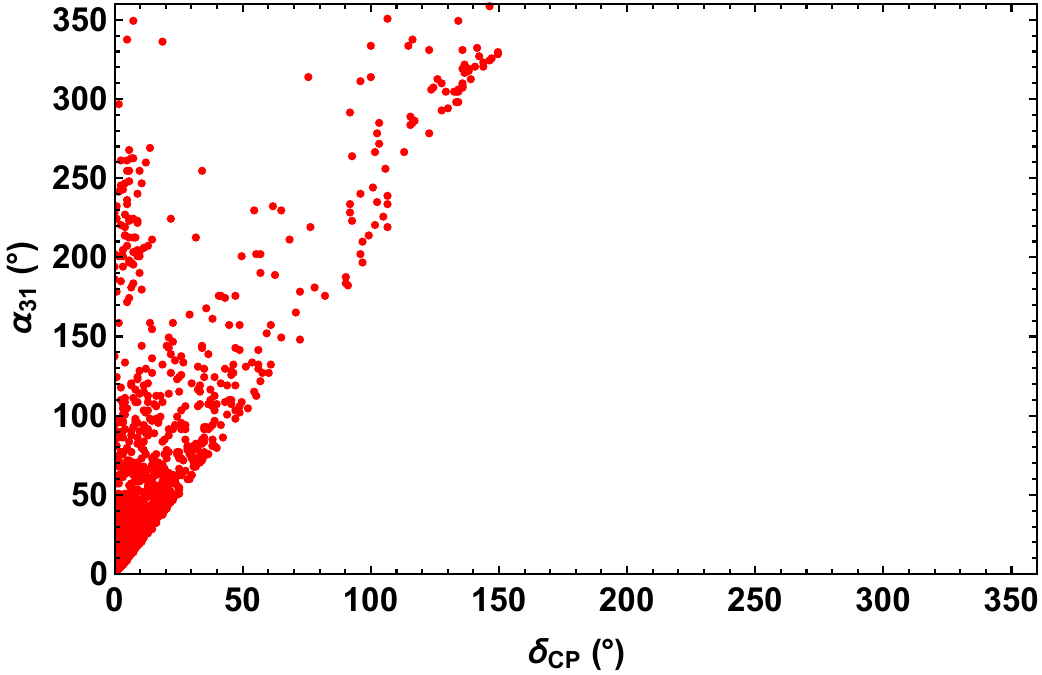} \\
	\end{tabular}
	\caption{Correlation between Majorana phases and Dirac CP phases for NO(left) and IO(right).}
	\label{fig:plot12}
 \end{figure}
 \par Also, we have calculated Majorana phases for the model. The figure \ref{fig:plot12} implies that $\alpha_{21}$ takes the value in the range $(0^\circ-360^\circ)$ for NO and $(0^\circ-280^\circ)$for IO. Similarly, the ranges for $\beta$ are found to be $(0^\circ-360^\circ)$ for both NO and IO. 
\par
\begin{figure}[H]
	\centering
	\begin{tabular}{cccc}
		\includegraphics[width=0.45\textwidth]{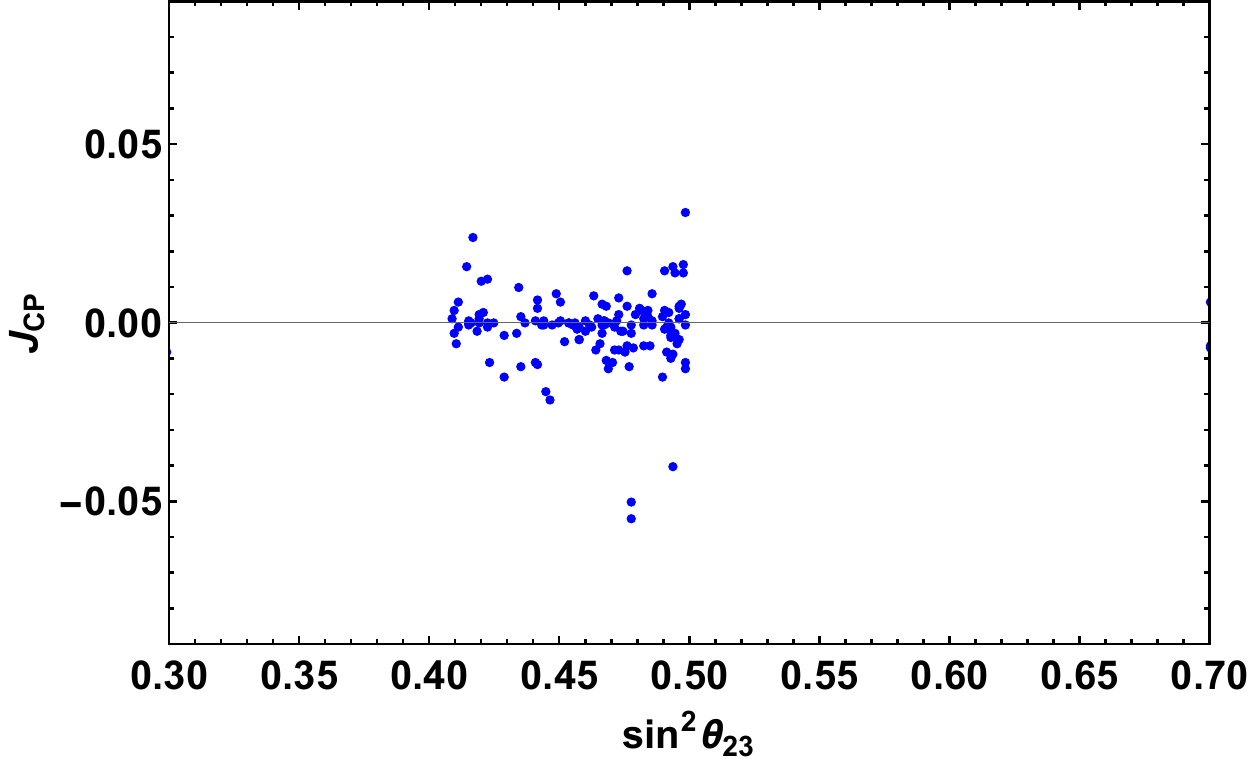} &
		\includegraphics[width=0.45\textwidth]{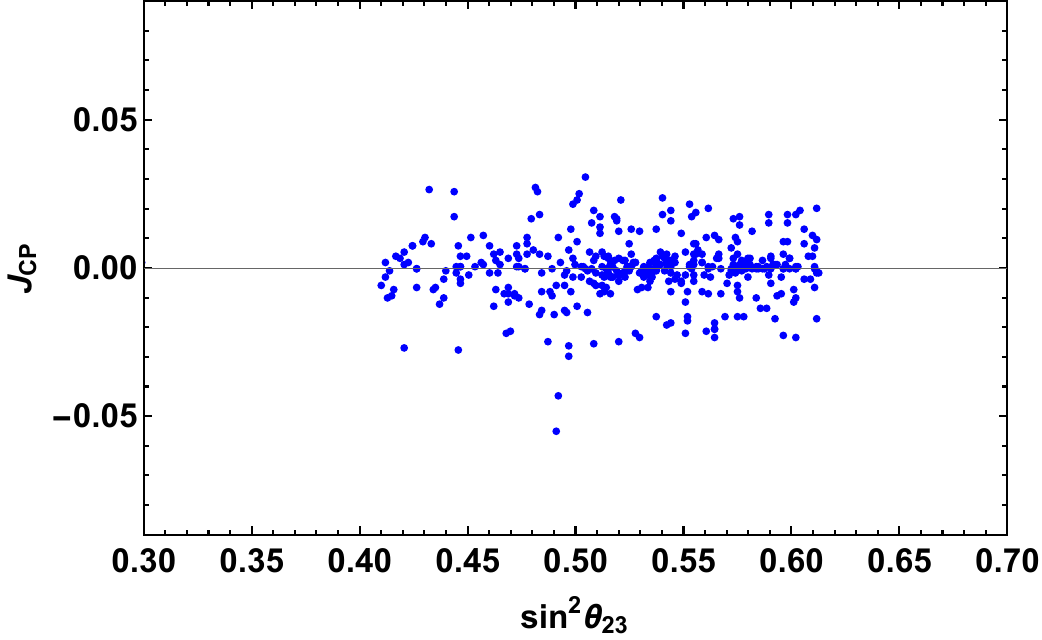} \\
	\end{tabular}
	\caption{Correlation between Jarsklog invariant $(J_{CP})$ with Dirac CP phase $(\delta_{CP})$ for NO (left) and IO (right).}
	\label{fig:plot13}
\end{figure}

 The correlation between the Dirac CP phase $\text{sin}^2\theta_{23}$ and the Jarsklog invariant $(J_{CP})$ is shown in figure \ref{fig:plot13}. The $J_{CP}$ is found to be in the region $-0.03$ to $0.03$ for both normal and inverted orderings.
\section{Lepton flavor violating processes} \label{sec_6}
\par In the framework of the present model, we will investigate the lepton flavor violation\cite{deppisch2013lepton} effects on $l_i \rightarrow l_j\gamma$. Lepton flavor violating decays usually forbidden in the Standard Model (SM), however they can be generated through the extension of lepton sector. Based on our model, the Majorana neutrino mass production mechanism can lead to charged LFV. Because $\frac{\Delta m_{21}^2}{m_{W}^2}\approx10^{-50}$ is well beyond the sensitivity reach of ongoing and future collider experiments, the GIM mechanism strictly suppresses LFV. However, in the case of a left-right symmetric model, charged LFV might arise as a result of the contribution from heavy right-handed neutrinos. According to the MEG collaboration\cite{meg2016search}, the current limit on the branching ratios is $\text{BR}(\mu\rightarrow e\gamma)<4.2\times10^{-13}$; for the BABAR collaboration\cite{aubert2010searches}, the limit is $\text{BR}(\tau\rightarrow \mu\gamma)<4.4\times10^{-8}$ with $90\%$ confidence level.
\par The purpose of COMET\cite{kurup2011coherent} and Mu2e\cite{kutschke2009mu2e}, among other planned experiments, is to achieve the conversion rate for the process $R^N(\mu\rightarrow e)$ in a nucleus\cite{cirigliano2004lepton} at a rate approximately $10^{-17}$.
\par The prediction of Lepton Flavor Violation is affected in our model by unitarity violation. The derivation of various constraints for the model relies significantly on $\mu\rightarrow e\gamma$, as the unitarity violation is of the order of $\frac{m_{D}^2}{M}$. According to \cite{ibarra2011low}, the branching ratio for the process $\mu\rightarrow e\gamma$ is
\begin{equation}
	\text{BR}(\mu\rightarrow e\gamma)=\frac{3\alpha}{32\pi}\sum_{i=1}^{3}f\biggl(\frac{M_i^2}{m_W^2}\biggr){|F_{\mu i}^*F_{ei}|}^2
\end{equation}
where $M_i$ denotes the physical masses for pseudo-Dirac particles.
\par
\begin{figure}[H]
	\centering
	\begin{tabular}{cccc}
		\includegraphics[width=0.45\textwidth]{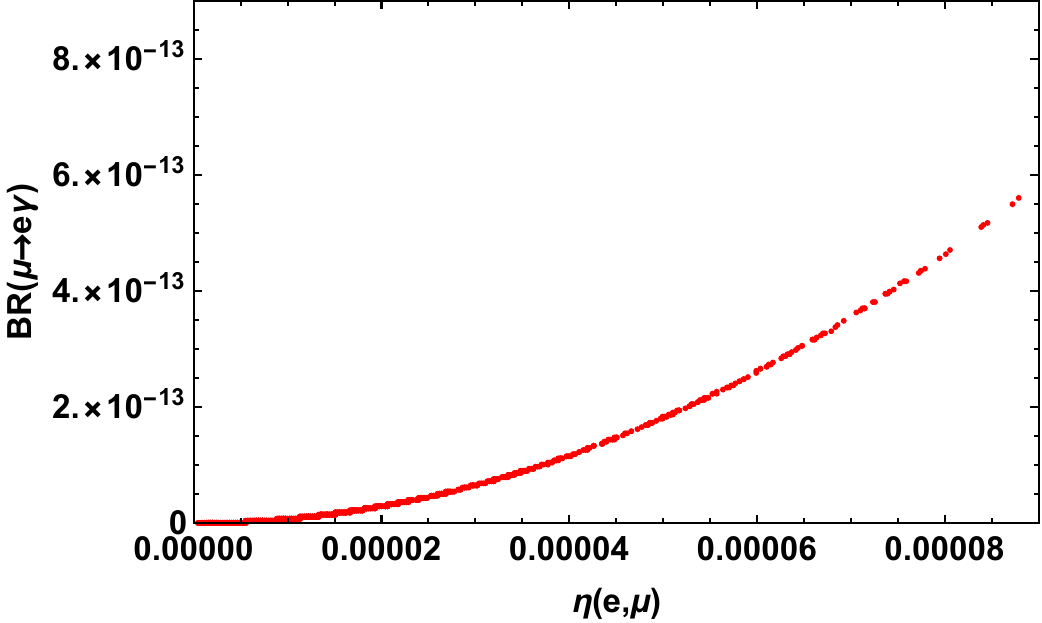} &
		\includegraphics[width=0.45\textwidth]{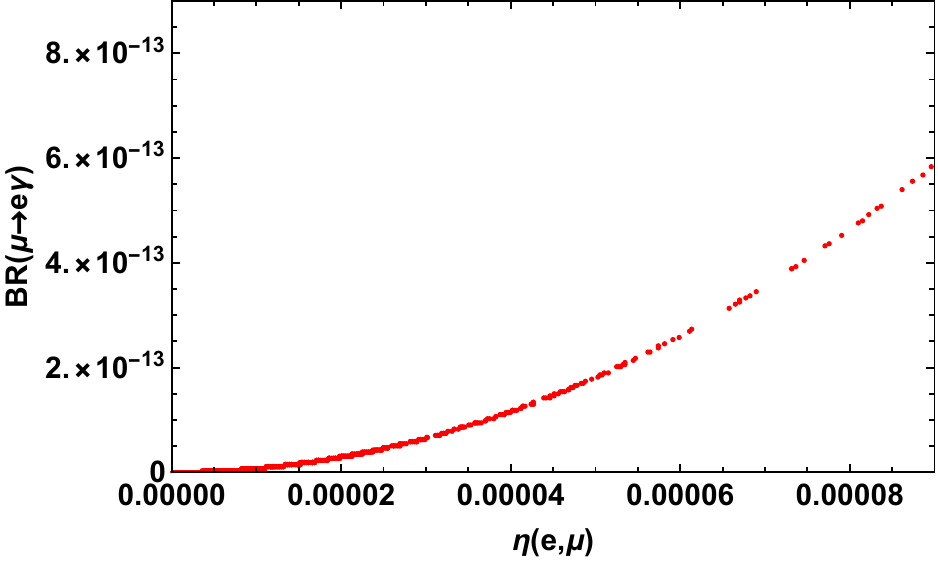} \\
	\end{tabular}
	\caption{Left and right panels represent the variation of $\text{BR}(\mu\rightarrow e\gamma)$ with $\eta(e,\mu)$ for NO(left) and IO(right).}
	\label{fig:plot14}
\end{figure}

 Figure \ref{fig:plot14} shows the Branching ratio for the process $\text{BR}(\mu\rightarrow e\gamma)$ for NO and IO.
\par The branching ratio for the process $\mu\rightarrow e\gamma$ is below $6\times10^{-13}$ for both NO and IO, which is within the current experimental reach, as shown in figures \ref{fig:plot14}.
\section{Leptogenesis in the present model} \label{sec_7}
\par The observed baryon asymmetry of the universe (BAU), which is determined to be $\eta_{B}^{\text{ obs}}=6.1\times10^{-10}$\cite{aghanim2020planck}, can be best explained through leptogenesis. It is closely related to the seesaw mechanism involving the violation of lepton number due to right-handed Majorana neutrinos. A very high right-handed neutrino mass of the order $10^{9-14}\text{ GeV}$\cite{davidson2002lower,dev2014flavour, branco2009resonant} is required for the standard thermal leptogenesis\cite{davidson2008leptogenesis} process, and this mass is beyond the scope of current collider experiments. However, this scale has been lowered to the TeV range via resonant leptogenesis\cite{pilaftsis2004resonant,pilaftsis1997cp,abada2019low}, where heavy neutrinos are quasi-degenerate and their mass splitting is of the order of their decay rates.
\par The heavy quasi-Dirac pairs in our model have a negligibly small mass splitting, which hinders the generation of the necessary lepton asymmetry. Therefore, the right-handed neutrinos need to have a small Majorana mass $M_R$ in order to generate the necessary mass splitting. Now, the Lagrangian for Majorana mass term can be obtained by
\begin{equation}
	\mathcal{L}_{R}=\alpha_{R}YSS\frac{H_R^{4}}{\Lambda^3}
\end{equation}
\par For the right-handed Majorana term, the mass matrix is constructed as 
\begin{equation}
	M_R=\frac{\alpha_R v_R^{4}}{3\Lambda^{3}}
	\begin{pmatrix}
		2Y_1 & -Y_3 & -Y_2\\
		-Y_3 & 2Y_2 & -Y_1\\
		-Y_2 & -Y_1 & 2Y_3\\
	\end{pmatrix}
\end{equation}
\par Now, the $2\times2$ submatrix of Eqn. (8) can be written in the basis $(\nu_R^{c},S)$ as
\begin{equation}
	\begin{pmatrix}
		0 & m_{RS} \\
		m_{RS}^T & M_R \\
	\end{pmatrix}
\end{equation}
\par The unitary matrix $\frac{1}{\sqrt{2}}\begin{pmatrix}
	I & -I \\
	I & I \\
\end{pmatrix}$ can be used to construct a block diagonal structure for the matrix given in Eqn. (33) as
\begin{equation}
	M^{'}=\begin{pmatrix}
		m_{RS}+\frac{M_R}{2} & -\frac{M_R}{2} \\
		-\frac{M_R}{2} & -m_{RS}+\frac{M_R}{2} \\
	\end{pmatrix} \approx
	\begin{pmatrix}
		m_{RS}+\frac{M_R}{2} & 0 \\
		0 & -m_{RS}+\frac{M_R}{2} \\
	\end{pmatrix}
\end{equation}
\par Thus, $N_R$ and $S$ are related to the mass eigenstates $(N_{\pm})$ by

\begin{equation*}
	\begin{pmatrix}
		S_{i} \\ 
		N_{Ri} \\
	\end{pmatrix}=\begin{pmatrix}
		\cos\theta & -\sin\theta \\
		\sin\theta & \cos\theta \\
	\end{pmatrix}
	\begin{pmatrix}
		N_{i}^{+} \\
		N_{i}^{-} \\
	\end{pmatrix}
\end{equation*}
\par Assuming maximal mixing, we can write 
\begin{equation*}
	N_{Ri}=\frac{N_i^{+}+N_i^{-}}{\sqrt{2}}, \; S_{i}=\frac{N_i^{+}-N_i^{-}}{\sqrt{2}}
\end{equation*}
\par As a result, the Lagrangian interaction can be modified and expressed in the new basis as
\begin{multline}
	\mathcal{L}_{LS}=\alpha_{LS}(\bar{l}_{L_e})_1H_L\Biggl[Y\Biggl(\frac{N_i^{+}-N_i^{-}}{\sqrt{2}}\Biggr)\Biggr]_1+\beta_{LS}(\bar{l}_{L_\mu})_{1^{''}}H_L\Biggl[Y\Biggl(\frac{N_i^{+}-N_i^{-}}{\sqrt{2}}\Biggr)\Biggr]_{1^{'}} \\
	+\gamma_{LS}(\bar{l}_{L_\tau})_{1^{'}}H_L\Biggl[Y\Biggl(\frac{N_i^{+}-N_i^{-}}{\sqrt{2}}\Biggr)\Biggr]_{1^{''}}
\end{multline}
\par Now, the modified mass matrix is given by 
\begin{equation}
	m_{RS}\pm\frac{M_R}{2}=\Biggl(\frac{\alpha_{RS}v_{RS}}{3}\pm\frac{\alpha_{R}v_{R}^{4}}{6\Lambda^{3}}\Biggr)\begin{pmatrix}
		2Y_1 & -Y_3 & -Y_2\\
		-Y_3 & 2Y_2 & -Y_1\\
		-Y_2 & -Y_1 & 2Y_3\\
	\end{pmatrix}
\end{equation}
\par Through $(M^{\pm})_{\text diag}=U_{\text TBM}U_{R}(M_{RS}\pm\frac{M_R}{2})U_{R}^{T}U_{\text TBM}^{T}$, the aforementioned matrix can be diagonalised and the mass eigenvalues are found to be
\begin{equation*}
	M_{1}^{\pm}=\frac{1}{2}\Biggl(\frac{\alpha_{RS}v_{RS}}{3}\pm\frac{\alpha_{R}v_{R}^{4}}{6\Lambda^{3}}\Biggr)\Biggl(Y_1+Y_2+Y_3+\sqrt{9Y_1^{2}-6Y_1Y_2+3Y_2^{2}-2Y_1Y_3+3Y_3^{2}}\Biggr)
\end{equation*}
\begin{equation*}
	M_{2}^{\pm}=\frac{1}{2}\Biggl(\frac{\alpha_{RS}v_{RS}}{3}\pm\frac{\alpha_{R}v_{R}^{4}}{6\Lambda^{3}}\Biggr)\Biggl(Y_1+Y_2+Y_3-\sqrt{9Y_1^{2}-6Y_1Y_2+3Y_2^{2}-2Y_1Y_3+3Y_3^{2}}\Biggr)
\end{equation*}
\begin{equation*}
	M_{3}^{\pm}=\frac{1}{2}\Biggl(\frac{\alpha_{RS}v_{RS}}{3}\pm\frac{\alpha_{R}v_{R}^{4}}{6\Lambda^{3}}\Biggr)(Y_1+Y_2+Y_3)
\end{equation*}
\par Now, the CP asymmetry can be expressed as\cite{pilaftsis1997cp,gu2010leptogenesis}
\begin{equation}
	\epsilon_{N_{i}^{-}}\approx\frac{1}{32\pi^{2}A_{N_{i}^{-}}}\text{Im}\Biggl[\Bigg(\frac{\tilde{m}_{D}}{v_u}-\frac{\tilde{m}_{LS}}{v_u}\Biggr)^\dagger\Bigg(\frac{\tilde{m}_{D}}{v_u}+\frac{\tilde{m}_{LS}}{v_u}\Biggr)^2\Bigg(\frac{\tilde{m}_{D}}{v_u}-\frac{\tilde{m}_{LS}}{v_u}\Biggr)^\dagger\Biggr]_{ii}\frac{r_N}{r_{N}^{2}+4A_{N_{i}^{-}}^{2}}
\end{equation}
\par where $\tilde{m}_{D}=m_{D}U_{TBM}U_R$, $\tilde{m}_{LS}=m_{LS}U_{TBM}U_R$ and $r_N$ and $A_N^{-}$ are given by
\begin{equation}
	r_N=\frac{(M_i^{+})^2-(M_i^{-})^2}{M_i^{+}M_i^{-}}=\frac{\Delta{M}(M_i^{+}+M_i^{-})}{M_i^{+}M_i^{-}},  A_{N^{-}}\approx\frac{1}{16\pi}\Biggl[\Biggl(\frac{\tilde{m}_{D}}{v_u}-\frac{\tilde{m}_{LS}}{v_u}\Biggr)\Biggl(\frac{\tilde{m}_{D}}{v_u}+\frac{\tilde{m}_{LS}}{v_u}\Biggr)\Biggr]_{ii}
\end{equation}
\par Here, $\Delta M=M_i^{+}-M_i^{-}\approx M_R$.
\par Because modular symmetry is imposed, eliminating extra flavon fields, CP asymmetry parameter of the model greatly relies on the Yukawa couplings in addition to other free parameters. The dynamics of Boltzmann equations describes the evolution of lepton asymmetry. Now, in order to implement the Sakharov conditions\cite{sakharov1998violation}, one requires
\begin{equation*}
	K=\frac{\Gamma_{N_{1}^{-}}}{H(T=M_1^{-})}
\end{equation*}
\par where $H=\frac{1.66\sqrt{g_*}T^2}{M_{Pl}}$ with $g_*=106.75$ and $M_{Pl}=1.22\times 10^{19}$ GeV.
\par Now, the Boltzmann equations are given by\cite{buchmuller2005leptogenesis,plumacher1997baryogenesis,giudice2004towards,strumia2006baryogenesis},
\begin{align}
	\frac{dY_N^{-}}{dz}=-\frac{z}{sH(M_1^{-})}\Biggl[\biggl(\frac{Y_{N^-}}{Y_{N^-}^{eq}}-1\biggr)\gamma_D+\Biggl(\biggl(\frac{Y_{N^-}}{Y_{N^-}^{eq}}\biggr)^2-1\Biggr)\gamma_s\Biggr], \nonumber \\
	\frac{dY_{B-L}}{dz}=-\frac{z}{sH(M_1^{-})}\Biggl[\epsilon_{N^-}\biggl(\frac{Y_{N^-}}{Y_{N^-}^{eq}}-1\biggr)\gamma_D-\frac{Y_{B-L}}{Y_l^{eq}}\frac{\gamma_D}{2}\Biggr],
\end{align}
where $s$ denotes the entropy density, $z=\frac{M_1^{-}}{T}$. Also, the equilibrium number densities can be written as\cite{davidson2008leptogenesis}
\begin{align}
	Y_{N^-}^{eq}&=\frac{45g_{N^-}}{4\pi^4g_*}z^2K_2(z), &  Y_l^{eq}&=\frac{3}{4}\frac{45\zeta(3)g_l}{2\pi^4g_*}
\end{align}
\par Here, $K_{1,2}$ are the modified Bessel functions, $g_l=2$ and $g_{N^-}=2$. The decay rate $\gamma_D$ can be expressed by the equation
\begin{equation*}
	\gamma_D=sY_{N^-}^{eq}\Gamma_D
\end{equation*}
where, $\Gamma_D=\Gamma_{N^-}\frac{K_1(z)}{K_2(z)}$. Moreover, $\gamma_s$ indicates scattering rate of the decaying particle $N_1^{-}N_1^{-}\rightarrow\rho\rho$\cite{iso2011resonant}.
\begin{figure}[H]
	\centering
	\begin{tabular}{cccc}
		\includegraphics[width=0.45\textwidth]{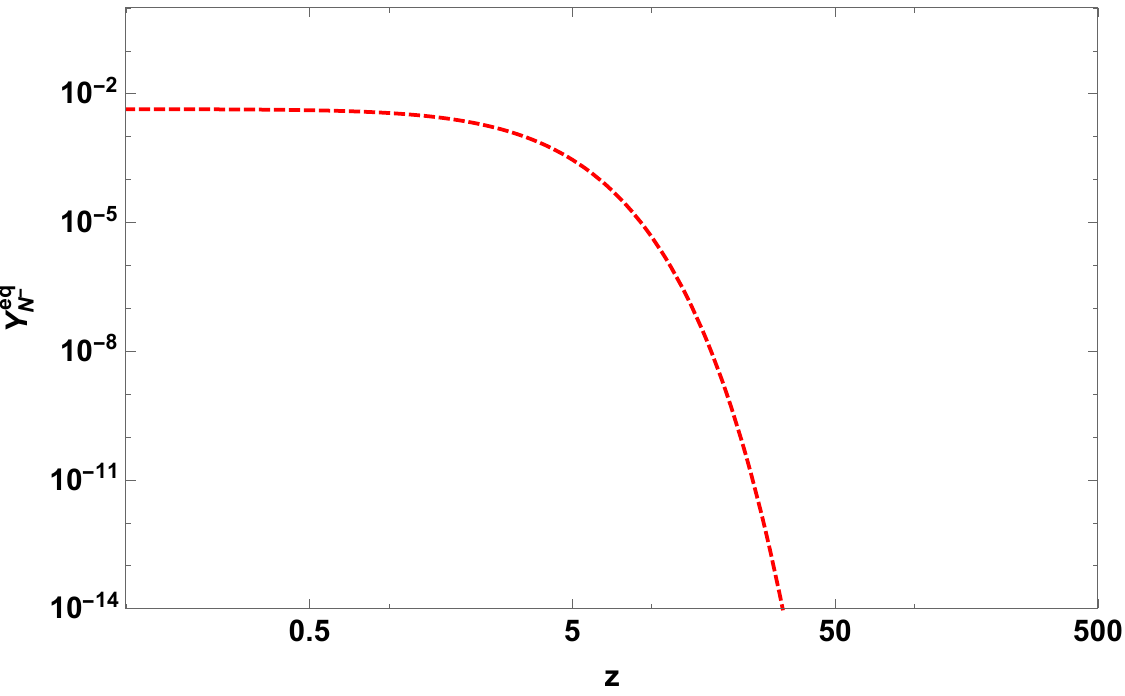} &
		\includegraphics[width=0.45\textwidth]{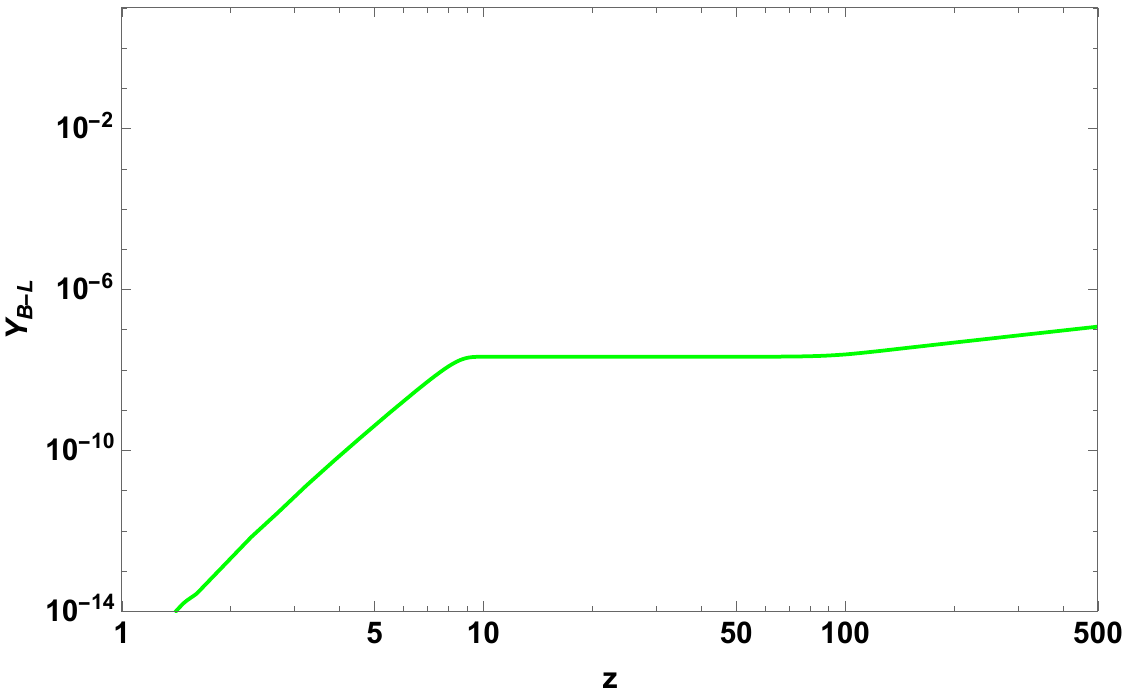} \\
	\end{tabular}
	\caption{Evolution of $Y_{N^-}^{eq}$(left) and $Y_{B-L}$(right) with $z$.}
	\label{fig:plot15}
\end{figure}
\section{Comment on Collider Physics}\label{sec_8}
\par Here, we briefly discuss the most promising collider implication of heavy pseudo-Dirac neutrinos within the present model that may be possible at the LHC, without providing any numerical calculations. Since the $M_{LS}$ is the term that violates the lepton number in the present scenario\cite{han2021full}, its mass scale is essentially tiny. Additionally, the effective Majorana neutrino mass matrix for active neutrinos appears in Eqn. \eqref{eqn_9} where the suppression of the pseudo-Dirac neutrino mass term, $M_{LS}$, is responsible for the smallness of $m_v$. This suppression is further attributed to the ratio of $M_D$ to $M_{RS}$. Also, trilepton plus missing energy is a significant mechanism involving heavy pseudo-Dirac neutrinos that may be studied at colliders\cite{chen2012multilepton,das2014direct}.

\begin{equation}
    \sigma(pp\rightarrow Nl^\pm\rightarrow l^\mp l^\pm+\cancel{E})=\sigma(pp\rightarrow W\rightarrow Nl^\pm)\times Br(N\rightarrow l^\mp l^\pm+\cancel{E})
\end{equation}
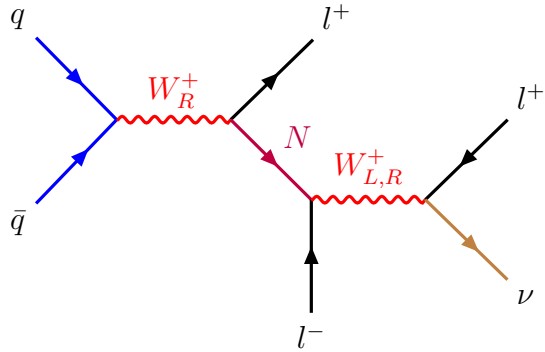
\begin{figure}[H]
\begin{center}
\begin{tikzpicture}
\begin{feynman}
    \vertex (a);
    \vertex [above left=of a] (b) {$q$};
    \vertex [below left=of a] (c) {$\bar{q}$};
    \vertex [right=of a] (d);
    \vertex [above right=of d] (e) {$l^+$};
    \vertex [below right=of d] (f);
    \vertex [below=of f] (g) {$l^-$};
    \vertex [right=of f] (h);
    \vertex [above right=of h] (i) {$l^+$};
    \vertex [below right=of h] (j) {$\nu$};

    \diagram{
    (b) -- [blue, fermion, very thick] (a) -- [blue, anti fermion, very thick] (c);
    (a) -- [red, boson, edge label=$W_R^{+}$, very thick] (d);
    (d) -- [black, fermion, very thick] (e);
    (d) -- [purple, fermion, edge label=$N$,very thick] (f);
    (g) -- [black, fermion, very thick] (f);
    (f) -- [red, boson, edge label=$W_{L,R}^{+}$, very thick] (h);
    (i) -- [black, fermion, very thick] (h) -- [brown, fermion, very thick]; (j)--[brown, anti fermion, very thick] (h);
    };
    \end{feynman}
\end{tikzpicture}
\caption{ The Feynman
diagram showing the decay process for $pp\rightarrow l^+l^-l^+\cancel{E}$ process}
	\label{fig:plot16}
\end{center}
\end{figure}
\par The possibility of producing this trilepton plus missing energy at the collider is mainly dependent on three factors: 
\par 1) a significant amount of mixing between heavy pseudo-Dirac neutrinos and light active neutrinos; 
\par 2) the mass of the heavy neutrinos, ideally ranging from a few [GeV-TeV]; 
\par 3) the method of production of this process.
\par The only approach available for differentiating between Majorana and pseudo-Dirac neutrinos at the collider is by thorough study of their decay processes. Similarly to the type-I seesaw mechanism, where heavy Majorana neutrinos at the TeV scale, the usual mixing between light-heavy neutrinos is $\theta_{\nu RS}\simeq\sqrt{m_\nu M_{RS}^{-1}}\leq10^{-6}$\cite{del2009distinguishing}.
\section{Conclusion} \label{sec_9}
\par In this study, we have examined a left-right symmetric model enhanced with $A_4$ modular flavor symmetry, in which the linear seesaw mechanism explains neutrino masses and mixing. This symmetry reduces the use of multiple flavon fields, improving the predictability of the model for both ongoing and upcoming collider experiments. Additionally, the neutrino mass matrix for the model, which is defined by the free parameters and modulus $\tau$, has been constructed.
\par After that, for both normal and inverted ordering cases, we have performed some numerical analysis to find parameter sets that can fit within the $3\sigma$ ranges of neutrino oscillation data. For both NO and IO, the lower bound for the sum of the neutrino masses is around $0.06$ eV. Also, the effective mass for neutrinoless double beta decay is found to be within the experimental bound for both NO and IO. Also, the Majorana phase $\alpha_{21}$ takes the value $(0^\circ-360^\circ)$ for NO and $(0^\circ-280^\circ)$ for IO. For Majorana phase $\alpha_{31}$, the range is $(0^\circ-360^\circ)$ for both NO and IO. Moreover, for both NO and IO, we have determined the $J_{CP}$ of order $10^{-2}$. Additionally, the linear seesaw mechanism has been our primary focus since it results in substantial light-heavy neutrino mixing, which is a significant contribution to lepton flavor violating decays such as $\mu\rightarrow e\gamma$, $\mu\rightarrow eee$, and $\mu\rightarrow e$. For the process $\mu\rightarrow e\gamma$, the branching ratio is found to be below $6\times10^{-13}$ for both NO and IO, which can be achieved by the ongoing and upcoming experiments.
\par We have solved coupled Boltzmann equations with the present set of model parameters to obtain the evolution of lepton asymmetry, which is found to be of the order of $\approx10^{-8}$. This is sufficient to explain the present baryon asymmetry of the universe. We have also demonstrated that the hierarchy of heavy neutrino mass is $M_1<<M_2<M_3$ for NO and $M_1<M_2\lesssim M_3$ for IO. Moreover, we have discussed about the implications of our model in collider physics, where the model may be tested in upcoming experiments.

\bibliographystyle{unsrt}
\bibliography{refs1}	

\begin{thebibliography}{1}

\bibitem{acciarri2016long}
R~Acciarri, MA~Acero, M~Adamowski, C~Adams, P~Adamson, S~Adhikari, Z~Ahmad,
  CH~Albright, T~Alion, E~Amador, et~al.,
\newblock Long-baseline neutrino facility {(LBNF)} and deep underground neutrino
  experiment {(DUNE)} conceptual design report, volume 4 the {DUNE} detectors at
  {LBNF},
\newblock {\em arXiv preprint arXiv:1601.02984}, 2016.

\bibitem{t2k2020constraint}
Constraint on the matter--antimatter symmetry-violating phase in neutrino
  oscillations.
\newblock {\em Nature}, 580(7803):339--344, 2020.

\bibitem{minkowski1977mu}
Peter Minkowski.
\newblock $\mu$→e$\gamma$ at a rate of one out of $10^9$ muon decays?
\newblock {\em Physics Letters B}, 67(4):421--428, 1977.

\bibitem{mohapatra1980neutrino}
Rabindra~N Mohapatra and Goran Senjanovi{\'c}.
\newblock Neutrino mass and spontaneous parity nonconservation.
\newblock {\em Physical Review Letters}, 44(14):912, 1980.

\bibitem{schechter1980neutrino}
Schechter, J and Valle, Jos{\'e} WF.
\newblock Neutrino masses in $SU(2) \bigotimes U(1)$ theories
\newblock {\em Physical Review D}, 22(9):2227, 1980.

\bibitem{zee1980theory}
A~Zee.
\newblock A theory of lepton number violation and neutrino {Majorana} masses.
\newblock {\em Physics Letters B}, 93(4):389--393, 1980.

\bibitem{babu1988model}
KS~Babu.
\newblock Model of “calculable” {Majorana} neutrino masses.
\newblock {\em Physics Letters B}, 203(1-2):132--136, 1988.

\bibitem{mohapatra2007theory}
Mohapatra, RN and Antusch, S and Babu, KS and Barenboim, Gabriela and Chen, Mu-Chun and De Gouv{\^e}a, A and De Holanda, P and Dutta, B and Grossman, Y and Joshipura, A and others.
\newblock Theory of neutrinos: a white paper.
\newblock {\em Reports on Progress in Physics}, 70(11):1757, 2007.

\bibitem{malinsky2005supersymmetric}
Malinsk{\`y}, M and Romao, JC and Valle, JWF.
\newblock Supersymmetric {SO(10)} seesaw mechanism with low {B-L} scale.
\newblock {\em Physical review letters}, 95(16):161801, 2005.

\bibitem{dib2014neutrinos}
Dib, Claudio O and Moreno, Gast{\'o}n R and Neill, Nicol{\'a}s A.
\newblock Neutrinos with a linear seesaw mechanism in a scenario of gauged {B-L} symmetry.
\newblock {\em Physical Review D}, 90(11):113003, 2014.

\bibitem{gonzalez1989fast}
M~Concepci{\'o}n Gonz{\'a}lez-Garci{\'a} and Jos{\'e}~WF Valle.
\newblock Fast decaying neutrinos and observable flavour violation in a new
  class of majoron models.
\newblock {\em Physics Letters B}, 216(3-4):360--366, 1989.

\bibitem{hernandez2019viable}
AE~C{\'a}rcamo Hern{\'a}ndez, Juan~Marchant Gonz{\'a}lez, and Ulises~Jesus
  Salda{\~n}a-Salazar.
\newblock Viable low-scale model with universal and inverse seesaw mechanisms.
\newblock {\em Physical Review D}, 100(3):035024, 2019.

\bibitem{brdar2019low}
Brdar, Vedran and Smirnov, Alexei Yu.
\newblock Low scale left-right symmetry and naturally small neutrino mass.
\newblock {\em Journal of High Energy Physics}, 2019(2):1--29, 2019.

\bibitem{brahmachari20084}
Brahmachari, Biswajoy and Choubey, Sandhya and Mitra, Manimala.
\newblock {$A_4$} flavor symmetry and neutrino phenomenology.
\newblock {\em Physical Review D}, 77(7):073008, 2008.

\bibitem{mukherjee2016neutrino}
Mukherjee, Ananya and Das, Mrinal Kumar.
\newblock Neutrino phenomenology and scalar Dark Matter with {$A_4$} flavor symmetry in Inverse and type II seesaw.
\newblock {\em Nuclear Physics B}, 913:643--663, 2016.

\bibitem{di2019neutrino}
Di Iura, Andrea and L{\'o}pez-Ib{\'a}{\~n}ez, ML and Meloni, Davide.
\newblock Neutrino masses and lepton mixing from {$A_5$} and {CP}.
\newblock {\em Nuclear Physics B}, 949:114794, 2019.

\bibitem{ma2006neutrino}
Ma, Ernest.
\newblock Neutrino mass matrix from {$S_4$} symmetry.
\newblock {\em Physics Letters B}, 632(2--3):352--356, 2006.

\bibitem{chakraborty2020predictive}
Chakraborty, Mainak and Krishnan, R and Ghosal, Ambar.
\newblock Predictive {$S_4$} flavon model with {TM1} mixing and baryogenesis through leptogenesis.
\newblock {\em Journal of High Energy Physics}, 2020(9):1--48, 2020.

\bibitem{pakvasa1978discrete}
Pakvasa, Sandip and Sugawara, Hirotaka.
\newblock Discrete symmetry and {Cabibbo} angle.
\newblock {\em Physics Letters B}, 73(1):61--64, 1978.

\bibitem{feruglio2019neutrino}
Feruglio, Ferruccio.
\newblock Are neutrino masses modular forms?
\newblock {\em From My Vast Repertoire… Guido Altarelli’s Legacy}, 227--266, 2019.

\bibitem{xing2020flavor}
Xing, Zhi-zhong.
\newblock Flavor structures of charged fermions and massive neutrinos.
\newblock {\em Physics Reports}, 854:1--147, 2020.

\bibitem{sahu2020$a_4$}
Sahu, Purushottam and Patra, Sudhanwa and Pritimita, Prativa.
\newblock {$A_4$} realization of left-right symmetric linear seesaw.
\newblock {\em arXiv preprint arXiv:2002.06846}, 2020.

\bibitem{behera2022implications}
Behera, Mitesh Kumar and Mishra, Subhasmita and Singirala, Shivaramakrishna and Mohanta, Rukmani.
\newblock Implications of {$A_4$} modular symmetry on neutrino mass, mixing and leptogenesis with linear seesaw.
\newblock {\em Physics of the Dark Universe}, 36:101027 2022.

\bibitem{nomura2022linear}
Nomura, Takaaki and Okada, Hiroshi.
\newblock A linear seesaw model with {$A_4$} modular flavor and local $U(1)_{B-L}$ symmetries.
\newblock {\em Journal of Cosmology and Astroparticle Physics}, 2022(09):049 2022.

\bibitem{grimus1993introduction}
Grimus, Walter.
\newblock Introduction to left-right symmetric models.
\newblock {\em Vienna Univ.(Austria). Inst. fuer Theoretische Physik}, 2022.

\bibitem{schechter1982neutrino}
Schechter, J and Valle, Jos{\'e} WF.
\newblock Neutrino decay and spontaneous violation of lepton number.
\newblock {\em Physical Review D}, 25(3):774 1982.

\bibitem{agostini2019probing}
Agostini, M and Bakalyarov, AM and Balata, M and Barabanov, I and Baudis, L and Bauer, C and Bellotti, E and Belogurov, S and Bettini, A and Bezrukov, L and others.
\newblock Probing {Majorana} neutrinos with double-$\beta$ decay.
\newblock {\em Science}, 365(6460):1445--1448 2019.

\bibitem{alduino2018first}
Alduino, Chris and Alessandria, F and Alfonso, K and Andreotti, E and Arnaboldi, C and Avignone III, FT and Azzolini, O and Balata, M and Bandac, I and Banks, TI and others.
\newblock First {Results} from {CUORE}: {A} {Search} for {Lepton} {Number} {Violation} via $0\nu\beta\beta$ {Decay} of {Te}-130.
\newblock {\em Physical review letters}, 120(13):132501 2018.

\bibitem{giuliani2019double}
Giuliani, A and Cadenas, JJ and Pascoli, S and Previtali, E and Saakyan, R and Sch{\"a}ffner, K and Schoenert, S.
\newblock Double beta decay {APPEC} committee report.
\newblock {\em arXiv preprint arXiv:1910.04688}, 2019.

\bibitem{gando2016search}
Gando, Azusa and Gando, Y and Hachiya, T and Hayashi, A and Hayashida, S and Ikeda, H and Inoue, K and Ishidoshiro, K and Karino, Y and Koga, M and others.
\newblock Search for {Majorana} neutrinos near the inverted mass hierarchy region with {KamLAND-Zen}.
\newblock {\em Physical review letters}, 117(8):082503 2016.

\bibitem{de2012finite}
de Adelhart Toorop, Reinier and Feruglio, Ferruccio and Hagedorn, Claudia.
\newblock Finite modular groups and lepton mixing.
\newblock {\em Nuclear Physics B}, 117(8):082503 2016.

\bibitem{kobayashi2018neutrino}
Kobayashi, Tatsuo and Tanaka, Kentaro and Tatsuishi, Takuya H.
\newblock Neutrino mixing from finite modular groups.
\newblock {\em Physical Review D}, 858(3):437--467, 2012.

\bibitem{kobayashi2018modular}
Kobayashi, Tatsuo and Nagamoto, Satoshi and Takada, Shintaro and Tamba, Shio and Tatsuishi, Takuya H.
\newblock Modular symmetry and non-Abelian discrete flavor symmetries in string compactification.
\newblock {\em Physical Review D}, 97(11):116002, 2018.

\bibitem{king2020fermion}
King, Simon JD and King, Stephen F.
\newblock Fermion mass hierarchies from modular symmetry.
\newblock {\em Journal of High Energy Physics}, 2020(9):1--30 2020.

\bibitem{forero2011lepton}
Forero, DV and Morisi, S and Tortola, M and Valle, JWF.
\newblock Lepton flavor violation and non-unitary lepton mixing in low-scale type-I seesaw.
\newblock {\em Journal of High Energy Physics}, 2011(9):1--18 2011.

\bibitem{fernandez2016global}
Fernandez-Martinez, Enrique and Hernandez-Garcia, Josu and Lopez-Pavon, Jacobo.
\newblock Global constraints on heavy neutrino mixing.
\newblock {\em Journal of High Energy Physics}, 2016(8):1--31 2016.

\bibitem{antusch2014non}
Antusch, Stefan and Fischer, Oliver.
\newblock Non-unitarity of the leptonic mixing matrix: {Present} bounds and future sensitivities.
\newblock {\em Journal of High Energy Physics}, 2014(10):1--30, 2014.

\bibitem{blennow2017non}
Blennow, Mattias and Coloma, Pilar and Fernandez-Martinez, Enrique and Hernandez-Garcia, Josu and Lopez-Pavon, Jacobo.
\newblock Non-unitarity, sterile neutrinos, and non-standard neutrino interactions.
\newblock {\em Journal of High Energy Physics}, 2017(4):1--26, 2017.

\bibitem{agostinho2018can}
Agostinho, Nuno Rosa and Branco, GC and Pereira, Pedro MF and Rebelo, MN and Silva-Marcos, JI.
\newblock Can one have significant deviations from leptonic 3 $\times$ 3 unitarity in the framework of type I seesaw mechanism?
\newblock {\em The European Physical Journal C}, 78:1--11 2018.

\bibitem{gonzalez2021nufit}
Gonzalez-Garcia, Maria Concepcion and Maltoni, Michele and Schwetz, Thomas.
\newblock {NuFIT}: three-flavour global analyses of neutrino oscillation experiments.
\newblock {\em Universe}, 7(12):459 2021.

\bibitem{deppisch2013lepton}
Deppisch, Frank F.
\newblock Lepton flavour violation and flavour symmetries.
\newblock {\em Fortschritte der Physik}, 61(4-5):622--644, 2013.

\bibitem{meg2016search}
MEG collaboration and others.
\newblock Search for the {Lepton} {Flavour} {Violating} {Decay} $\mu^{+}\rightarrow e^{+}\gamma$ with the {Full} {Dataset} of the {MEG} {Experiment}.
\newblock {\em arXiv preprint arXiv:1605.05081}, 2016.

\bibitem{aubert2010searches}
Aubert, Bernard and Karyotakis, Y and Lees, JP and Poireau, V and Prencipe, E and Prudent, X and Tisserand, V and Tico, J Garra and Grauges, E and Martinelli, M and others.
\newblock Searches for lepton flavor violation in the decays $\tau^{\pm}\rightarrow e^{\pm}\gamma$ and $\tau^{\pm}\rightarrow \mu^{\pm}\gamma$.
\newblock {\em Physical review letters}, 104(2):021802, 2010.

\bibitem{kurup2011coherent}
Kurup, Ajit.
\newblock The coherent muon to electron transition ({COMET}) experiment.
\newblock {\em Nuclear Physics B-Proceedings Supplements}, 218(1):38--43, 2011.

\bibitem{kutschke2009mu2e}
Kutschke, Robert K.
\newblock The Mu2e Experiment at Fermilab.
\newblock{\em AIP Conference Proceedings}, 1182(1):718--721, 2009.

\bibitem{cirigliano2004lepton}
Cirigliano, V and Kurylov, A and Ramsey-Musolf, MJ and Vogel, P.
\newblock Lepton flavor violation without supersymmetry.
\newblock {\em Physical Review D—Particles, Fields, Gravitation, and Cosmology}, 70(7):075007, 2004.

\bibitem{ibarra2011low}
Ibarra, Alejandro and Molinaro, E and Petcov, ST.
\newblock Low energy signatures of the {TeV} scale seesaw mechanism.
\newblock{\em Physical Review D}, 84(1):013005, 2011.

\bibitem{aghanim2020planck}
Aghanim, Nabila and Akrami, Yashar and Ashdown, Mark and Aumont, Jonathan and Baccigalupi, Carlo and Ballardini, Mario and Banday, Anthony J and Barreiro, RB and Bartolo, Nicola and Basak, S and others.
\newblock Planck 2018 results-{$VI$}. {Cosmological} parameters.
\newblock {\em Astronomy \& Astrophysics}, 641:A6, 2020.

\bibitem{davidson2002lower}
Davidson, Sacha and Ibarra, Alejandro.
\newblock A lower bound on the right-handed neutrino mass from leptogenesis.
\newblock {\em Physics Letters B}, 535(1--4):105--177, 2002.

\bibitem{dev2014flavour}
Dev, PS Bhupal and Millington, Peter and Pilaftsis, Apostolos and Teresi, Daniele.
\newblock Flavour covariant transport equations: an application to resonant leptogenesis.
\newblock {\em Nuclear Physics B}, 886:569--664, 2014.

\bibitem{branco2009resonant}
Branco, GC and Felipe, R Gonzalez and Rebelo, MN and Serodio, H.
\newblock Resonant leptogenesis and tribimaximal leptonic mixing with {$A_4$} symmetry.
\newblock {\em Physical Review D}, 79(9):093008, 2009.

\bibitem{davidson2008leptogenesis}
Davidson, Sacha and Nardi, Enrico and Nir, Yosef.
\newblock Leptogenesis.
\newblock {\em Physics Reports}, 466(4--5):25--32, 2008.

\bibitem{pilaftsis2004resonant}
Pilaftsis, Apostolos and Underwood, Thomas EJ.
\newblock Resonant leptogenesis.
\newblock {\em Nuclear Physics B}, 692(3):303--345, 2004.

\bibitem{pilaftsis1997cp}
Pilaftsis, Apostolos.
\newblock {CP} violation and baryogenesis due to heavy {Majorana} neutrinos.
\newblock {\em Physical Review D}, 56(9):5431, 1997.

\bibitem{abada2019low}
Abada, Asmaa and Arcadi, Giorgio and Domcke, Valerie and Drewes, Marco and Klaric, Juraj and Lucente, Michele.
\newblock Low-scale leptogenesis with three heavy neutrinos.
\newblock {\em Journal of High Energy Physics}, 2019(1):1--54, 2019.

\bibitem{gu2010leptogenesis}
Gu, Pei-Hong and Sarkar, Utpal.
\newblock Leptogenesis with linear, inverse or double seesaw.
\newblock {\em Physics Letters B}, 694(3):226--232, 2010.

\bibitem{sakharov1998violation}
Sakharov, Andrei D.
\newblock Violation of {CP}-invariance, {C}-asymmetry, and baryon asymmetry of the {Universe}.
\newblock {\em In The Intermissions… Collected Works on Research into the Essentials of Theoretical Physics in Russian Federal Nuclear Center, Arzamas-16}, 694(3):84--87, 1998.

\bibitem{buchmuller2005leptogenesis}
Buchm{\"u}ller, Wilfried and Di Bari, Pasquale and Pl{\"u}macher, Michael.
\newblock Leptogenesis for pedestrians.
\newblock {\em Annals of Physics}, 315(2):305--351, 2005.

\bibitem{plumacher1997baryogenesis}
Pl{\"u}macher, Michael.
\newblock Baryogenesis and lepton number violation.
\newblock {\em Zeitschrift f{\"u}r Physik C Particles and Fields}, 74:(549--559), 1997.

\bibitem{giudice2004towards}
Giudice, Gian Francesco and Notari, A and Raidal, M and Riotto, A and Strumia, A.
\newblock Towards a complete theory of thermal leptogenesis in the {SM} and {MSSM}.
\newblock {\em Nuclear Physics B}, 685(1-3):89--149, 2004.

\bibitem{strumia2006baryogenesis}
Strumia, Alessandro.
\newblock Baryogenesis via leptogenesis.
\newblock {\em Les Houches}, 84:655--680, 2006.
\newblock {\em Physical Review D}, 98(1):016004, 2018.

\bibitem{iso2011resonant}
Iso, Satoshi and Okada, Nobuchika and Orikasa, Yuta.
\newblock Resonant leptogenesis in the minimal {B-L} extended standard model at {TeV}.
\newblock {\em Physical Review D}, 83(9):093011, 2011.

\bibitem{han2021full}
Han, He-chong and Xing, Zhi-zhong.
\newblock A full parametrization of the 9$\times$ 9 active-sterile flavor mixing matrix in the inverse or linear seesaw scenario of massive neutrinos.
\newblock {\em Nuclear Physics B}, 973:115609, 2021.

\bibitem{chen2012multilepton}
Chen, Chien-Yi and Dev, PS Bhupal.
\newblock Multilepton collider signatures of heavy {Dirac} and {Majorana} neutrinos.
\newblock {\em Physical Review D—Particles, Fields, Gravitation, and Cosmology}, 85(9):093018, 2012.

\bibitem{das2014direct}
Das, Arindam and Dev, PS Bhupal and Okada, Nobuchika.
\newblock Direct bounds on electroweak scale pseudo-Dirac neutrinos from $s = 8$ {TeV} {LHC} data.
\newblock {\em Physics Letters B}, 735:364--370, 2014.

\bibitem{del2009distinguishing}
Del Aguila, F and Aguilar-Saavedra, JA.
\newblock Distinguishing seesaw models at {LHC} with multi-lepton signals.
\newblock {\em Nuclear Physics B}, 813(1-2):22--90, 2009.

\end{thebibliography}

\end{document}